\documentclass[11pt,a4paper]{article}
\usepackage{jcappub}
\usepackage[utf8]{inputenc}
\pdfoutput=1 
\usepackage{graphicx}
\usepackage[figuresright]{rotating}
\usepackage{amssymb}
\usepackage{bm}
\usepackage{color}
\usepackage{placeins}
 \usepackage{textcomp}

\usepackage{float}

\usepackage{booktabs} 
\usepackage{tablefootnote}
\usepackage{graphicx}
\usepackage{caption}
\usepackage{subcaption}
\usepackage{array}
\usepackage[shortlabels]{enumitem}

\definecolor{dgreen}{rgb}{0,0.6,0.0}

\newcommand{\de}{\delta}
\newcommand{\De}{\Delta}

\newcommand{\La}{\Lambda}
\newcommand{\Om}{\Omega}

\newcommand{\si}{\sigma}

\newcommand{\be}{\begin{equation}}
\newcommand{\ee}{\end{equation}}
\newcommand{\gsim}{\stackrel{>}{\sim}}

\newcommand{\bea}{\begin{eqnarray}}
\newcommand{\eea}{\end{eqnarray}}
\newcommand{\bean}{\begin{eqnarray*}}
\newcommand{\eean}{\end{eqnarray*}}

\newcommand{\id}{{\rm 1\kern -2.5pt I}} 
\newcommand{\bn}{{\mathbf n}}
\newcommand{\bv}{{\mathbf v}}

\newcommand{\cd}{{\cdot}}

\newcommand{\Omm}{\Omega_{{\rm m}}}

\begin{document}
\title{A local infall from a cosmographic analysis of Pantheon+}

\author[]{Francesco Sorrenti,}
\author[]{Ruth Durrer and}
\author[]{Martin Kunz}
\affiliation[]{D\'epartement de Physique Th\'eorique and Center for Astroparticle Physics,\\
Universit\'e de Gen\`eve, 24 quai Ernest  Ansermet, 1211 Gen\`eve 4, Switzerland}
\date{today}

\emailAdd{francesco.sorrenti@unige.ch}
\emailAdd{ruth.durrer@unige.ch}
\emailAdd{martin.kunz@unige.ch}

\abstract{We present a model independent analysis of the Pantheon+ supernova sample and study the  dependence of the recovered values of $H_0$, $q_0$ and $j_0$ on the redshift cut and on the modeling of peculiar velocities. In addition to the bulk velocity discussed previously, we also find a significant infall that we attribute to the presence of an overdensity out to a radius of $R\simeq 120h^{-1}$Mpc.
}

\date{April 2024}

\maketitle

\section{Introduction}
The expansion history of the Universe is one of the major ingredients of the standard model of cosmology. In a Friedmann universe, the Hubble function, $H(z)$, determines the only dynamical degree of freedom of the geometry. We usually determine it by measuring the distance out to some (normalizable) standard candles or standard rulers as a function of redshift. Typically we have many measurements within given redshift bins from candles/rulers in different directions $\bn$ on the sky. But in the real, fluctuating Universe, the distance redshift relation also depends on direction due to peculiar velocities of the source and of the observer and to perturbations in the metric. An expression of $d_L(z,\bn)$ within first order perturbation theory is given in~\cite{Bonvin:2005ps}.

At first order, the expectation value of the angular mean, the monopole, is not affected by the angular fluctuations, but the variance can be significant, depending on the redshift. In a typical realisation which our Universe might be, we expect deviations from the pure Friedmann universe $d_L(z)$ of the order of the standard deviation. It is well known that at low redshift, $z<0.5$ peculiar velocities are the dominant perturbations of the distance redshift relation. In this work we therefore concentrate on them. 

We use  Pantheon+, the largest publicly available supernovae type Ia (SNIa) dataset. The original Pantheon+ analysis~\cite{Brout:2022vxf} includes peculiar velocities by modelling them using galaxy surveys.
This model, however does not include a curl component of the velocity field and has considerable uncertainties. For these reasons we
prefer a more agnostic approach.

In previous works~\cite{Sorrenti_2023,Sorrenti:2024} we have found that peculiar velocities contribute a significant monopole, dipole and quadrupole to the Pantheon+ supernovae. We assumed a $\La$CDM cosmology and modelled the velocity redshift dependence by linear perturbation theory. Furthermore, we have analysed the data by excluding 
all supernovae below a certain redshift cut, to test how far out we can detect a dipole in the supernova observations by removing the nearby data points. In the present analysis we shall be more agnostic both on the redshift dependence of the velocity field and on the cosmological model. Also, we shall analyse the SNIa's {\em inside} spheres of a given redshift limit to study the local environment. 

Surprisingly, we find that within a ball centered at the observer and extending to a redshift of $z\simeq 0.04$, corresponding to a radius $R(z)=z/H_0 \simeq 120h^{-1}$Mpc, the velocities are infalling, hinting to an {\em overdensity} inside this sphere. Neglecting this infall leads to a smaller value of the Hubble constant. For a redshift limit $z^{\rm (lim)}=0.04$, i.e. considering supernovae with redshift $z\leq0.04$, this value agrees -- probably accidentally -- with the Cosmic Microwave Background (CMB) inferred value of $H_0$.

The paper is structured as follows. In the next section we present the properties of the Pantheon+ data most relevant for our analysis and we briefly describe our Markov Chain Monte Carlo (MCMC) method. In Section~\ref{s:ba} we present our agnostic distance ansatz and present results for redshift with respect to the rest frame of the CMB, $z_{\rm cmb}$. We also compare the cosmological parameters obtained for different redshift models.
In Section~\ref{s:res} we include monopole and dipole perturbations which peculiar velocities can induce. In Section~\ref{s:con} we discuss our results and conclude.

\section{The Pantheon+ dataset}
The Pantheon+ dataset, see~\cite{Brout:2022vxf}, consists of 1701  lightcurves from 1550 supernovae type Ia.
{For our analysis we use the publicly available data, covariances and codes provided by the Pantheon+ collaboration at \url{https://github.com/PantheonPlusSH0ES/DataRelease}. In particular, we run an MCMC maximizing the likelihood:
\be \label{eq:likelihood}
\log(\mathcal{L}) = - \frac{1}{2} \Delta \boldsymbol{\mu}^T C^{-1} \Delta \boldsymbol{\mu},
\ee
where $C$ is the covariance matrix and $\Delta \boldsymbol{\mu}$ is given by:
\be \label{e:cases}
\Delta \mu^i =
\begin{cases}
\mu^i + dM -\mu_\mathrm{ceph}^i,  \quad i \in  \text{Cepheid hosts} \, ,\\
\mu^i +  dM -\mu_\mathrm{model}^i, \quad \text{otherwise} \, .\\
\end{cases}
\ee 
Here $\mu^i$ is the distance modulus measured with the lightcurve $i$ and
$dM$ is a nuisance parameter that models the calibration uncertainties of the supernovae with Cepheids in galaxies that host both a supernova and at least one Cepheid star. The distance modulus is related to the luminosity distance $d_L$ via the relation
\be
\mu^i_\mathrm{model} = 5 \log \biggl( \frac{d_L(z_i, \bn_i)}{\rm Mpc} \biggr) + 25 \, .
\ee
The redshift $z_i$ in this expression is the theoretical redshift in the isotropic and homogeneous `background' universe. In reality, we have to use an observed redshift, which will be affected by Doppler shifts due to the motion of the observer or/and the source. The measured redshift can be corrected by including some of these shifts. This will in general change the results, something that we will study below.

Our MCMC routine is performed with \texttt{emcee}~\citep{emcee} and parallelized using \texttt{schwimmbad}~\citep{schwimmbad}. In all the following analyses we consider wide flat priors. The chain convergence is achieved after a number of iteration $N > 50 \tau_{\rm max}$, where $\tau_{\rm max}$ is the longest integrated autocorrelation time $\tau$~\cite{autocorr} computed for all the parameters. Finally, after discarding $2 \, \lfloor \tau_{\rm max} \rfloor$ burn-in steps from the chain, we generate the contour plots of our fits using the \texttt{getdist} library~\cite{getdist}. 
}

\begin{minipage}{0.35\linewidth}

\begin{table}[H]
\centering
\begin{tabular}{ c c }
\toprule
 $z^{\rm (lim)}$ & Number of SNIa\\
 \midrule
 0.02  & 272 \\
  0.03 & 466\\
 0.04  & 594 \\ 
 0.05  & 648\\
  0.06  & 679 \\
  0.07  & 703 \\
  0.08  & 727 \\
  0.09  & 738 \\
  0.1  & 741 \\
  0.2  & 949 \\
  0.3  & 1207 \\
  0.4  & 1393\\
  0.5  & 1493\\
  0.6  & 1573 \\
  0.7  & 1626 \\
  0.8  & 1671  \\
 \bottomrule
 \end{tabular}
 \vspace{1.1 cm}
 \caption{\small Number of SNe lightcurves for each data set as a function of the upper redshift limit. In our analysis we always consider all 77 Cepheids for all redshift cuts. \label{t:nSn}  }
 \end{table}

\end{minipage}~~\hspace{0.15 cm}
\begin{minipage}{0.55\linewidth}
\begin{figure}[H]
\begin{subfigure}{.5\textwidth}
  \centering
  \includegraphics[scale=0.5]{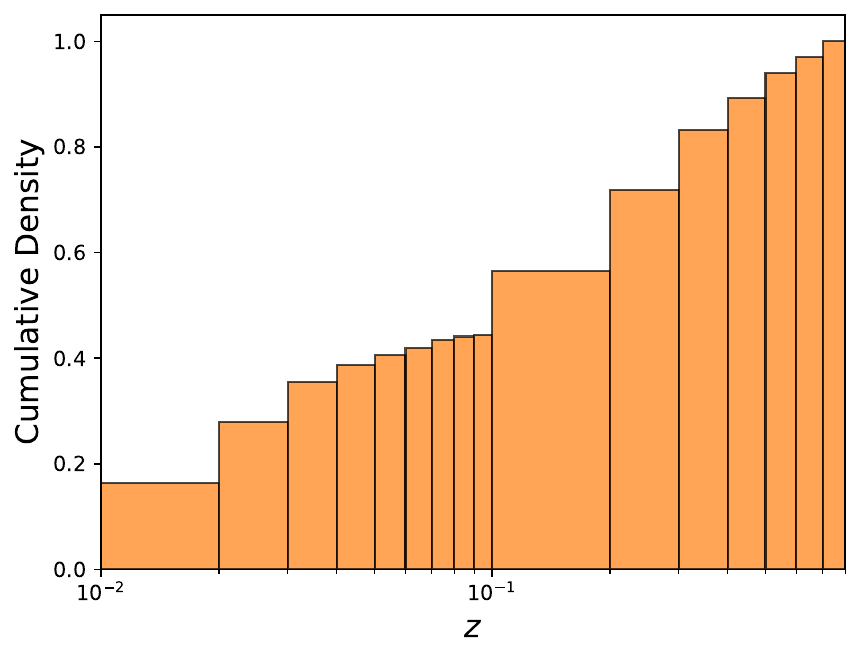}  
\end{subfigure}
\\
\begin{subfigure}{.5\textwidth}
  \centering
  \includegraphics[scale=0.5]{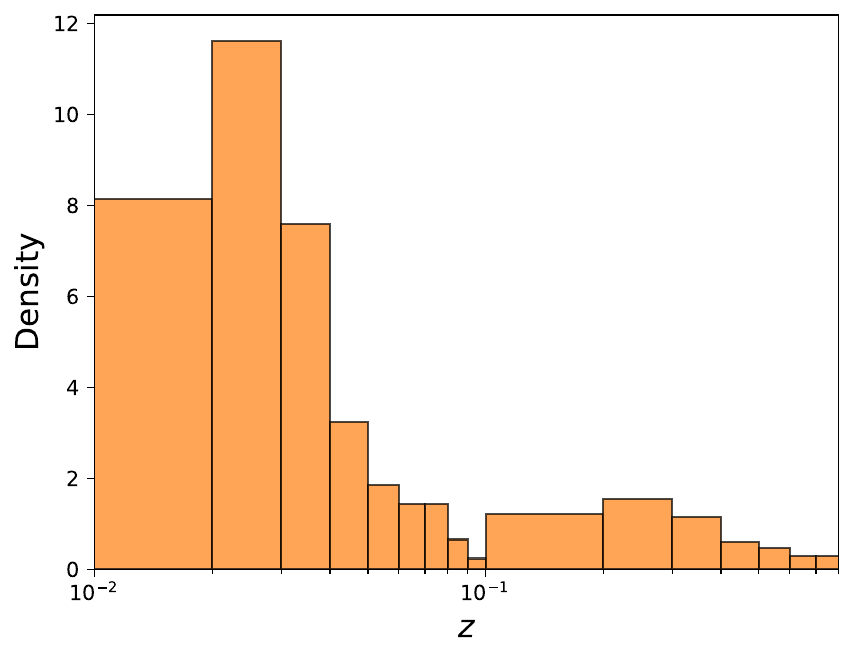}  
\end{subfigure}
\caption{\small Probability density for the redshift distributions (cumulative in the upper panel) of the SNe with $z \leq 0.8$ for the redshift bins that we use. \label{f:z_distributions}}
\vspace{0.2cm}
\end{figure}

\end{minipage}

In Table~\ref{t:nSn} and Fig.~\ref{f:z_distributions} we present the number of SNIa lightcurves in the data below a certain redshift $z$. Since we shall use a model agnostic Taylor expansion in $z$ for the luminosity distance we do not consider redshifts larger than $z=0.8$. As we see from the last line of Table~\ref{t:nSn}, this means that we do not include the 30 SNIa lightcurves with highest redshifts. Note also that more than 43\% of the Pantheon+ SNIa have redshifts below z=0.1. These are particularly important in our analysis.

\section{Agnostic distance measurements }\label{s:ba}

\subsection{Taylor expansion}
In this analayis we do not want to fit the parameters of a given cosmological model, but we want to remain as agnostic as possible. 
We simply expand the luminosity distance as a function of redshift in a Taylor series up to third order

\be \label{eq:agnostic:taylor_expansion_cmb}
d_L(z)=  d^{(1)}z + \frac{1}{2}  d^{(2)}z^2 + \frac{1}{6}  d^{(3)}z^3  \,,
\ee
where $d^{(n)}$ denotes the nth derivative of $d_L(z)$ at $z=0$.
The Taylor coefficients $d^{(n)}$, are related to the more standard parameters, the present expansion rate, $H_0=(\dot a/a)(t_0)$, the deceleration parameter, $q_0= -(\ddot a/a)(t_0)/H_0^2$, and the 'jerk', $j_0= (\dddot a/a)(t_0)/H_0^3$, by 
\begin{align}
    H_0=&\frac{1}{d^{(1)}} \label{e:H0} \\
    q_0=&1-\frac{d^{(2)}}{d^{(1)}} \qquad \left[=\frac 32 \Om_m-1 \right] \label{e:q0}\\
    j_0=&-1-\frac{d^{(3)}}{d^{(1)}}+\biggl(1-\frac{d^{(2)}}{d^{(1)}}\biggr)\biggl(4-3\frac{d^{(2)}}{d^{(1)}} \biggr) \qquad \big[=1 \big]\,.
    \label{e:j0}
\end{align}
The expressions in brackets above give the values of $q_0$ and $j_0$ for a flat $\La$CDM model for a given matter density parameter $\Om_m$. Note that in flat $\La$CDM, $j_0\equiv 1$.

For the standard $\La$CDM model with Planck parameters, the 3rd order approximation \eqref{eq:agnostic:taylor_expansion_cmb} has an error of 1.3\% at redshift $z=0.8$ and of 2.7\% at $z=1$. To be on the safe side, we only consider SNIa with $z\leq 0.8$.

Inserting the ansatz \eqref{eq:agnostic:taylor_expansion_cmb} into our MCMC analysis of the Pantheon+ data 
using CMB-Planck corrected redshifts for the proper motion of the observer, we obtain the values reported in Table~\ref{tab:agnostic_taylor_z_cmb} for $H_0$, $q_0$ and $j_0$ for different redshift cuts. Note that in this first analysis we do not include any peculiar velocities of the source. This strongly affects the inferred value of $H_0$ from very low redshift supernovae and we recover the value reported by the Pantheon+ collaboration only at $z^{\rm (lim)}\geq0.04$ (within $1\si$ errors). In Fig.~\ref{fig:z0_behaviour} we show the interesting behavior of $H_0$ as a function of the cutoff redshift. This shows how important the treatment of peculiar velocities is, especially for low redshifts. Not surprisingly, the deceleration and jerk parameters, which are only determined by the second and third term of the Taylor series, cannot be determined at low redshift. While $q_0$ is well determined for redshift limits $z^{\rm (lim)}\geq0.6$, the error of the jerk parameters remains large out to our largest considered redshift upper limit $z^{\rm (lim)}=0.8$.\\

\hspace{-0.9 cm}\begin{minipage}{0.485\linewidth}
\begin{table}[H]
\footnotesize
\centering
\setlength\tabcolsep{2.9pt} 
\begin{tabular}{cccc}
        \toprule
	$z^{(\rm cmb)}$ &$H_0$& $q_0$ & $j_0$ \\
	&\scriptsize{[km/s/Mpc]}& & \\

  \midrule
        0.02 & $61.0^{+3.3}_{-4.5}$ & $52 \pm 20$ & $3854^{+2000}_{-6000}$ \vspace{3 pt} \\
        0.03 & $66.3\pm 2.6$ & $14.3^{+7.7}_{-6.9}$ & $-77^{+130}_{-270}$ \vspace{3 pt} \\
        0.04 & $65.5\pm 2.0$ & $17.2 \pm 4.2$ & $-151^{+120}_{-240}$\vspace{3 pt} \\
        0.05 & $70.8\pm 1.8$ & $1.8 \pm 3.1$ & $-35^{+73}_{-140}$\vspace{3 pt} \\
        0.06 & $70.9\pm 1.7$ & $1.5 \pm 2.5$ & $-29^{+55}_{-100}$\vspace{3 pt} \\
        0.07 & $71.3\pm 1.5$ & $0.6\pm 2.0$ & $7^{+49}_{-83}$\vspace{3 pt} \\
        0.08 & $70.9\pm 1.4$ & $1.4 \pm 1.5$ & $-31^{+27}_{-46}$\vspace{3 pt} \\
        0.09 & $71.1\pm 1.4$ & $1.2\pm 1.4$ & $-26^{+24}_{-41}$ \vspace{3 pt} \\
        0.10 & $70.7\pm 1.3$ & $1.9\pm 1.3$ & $-43^{+18}_{-31}$ \vspace{3 pt} \\
        0.20 & $71.5\pm 1.1$ & $0.49\pm 0.43$ & $-8.6^{+3.5}_{-5.2}$ \vspace{3 pt} \\
        0.30 & $72.0\pm 1.0$ & $0.04\pm 0.23$ & $-3.1^{+1.7}_{-2.2}$ \vspace{3 pt} \\
        0.40 & $72.2\pm 1.0$ & $-0.10\pm 0.17$ & $-1.9^{+1.1}_{-1.3}$ \vspace{3 pt} \\
        0.50 & $72.4\pm 1.0$ & $-0.22\pm 0.13$ & $-0.92^{+0.79}_{-0.95}$ \vspace{3 pt} \\
        0.60 & $72.6\pm 1.0$ & $-0.36\pm 0.11$ & $0.28^{+0.66}_{-0.76}$ \vspace{3 pt} \\
        0.70 & $72.6\pm 1.0$ & 
        $-0.363\pm 0.092$& $0.24^{+0.52}_{-0.59}$ \vspace{3 pt} \\
        0.80 & $72.7\pm 1.0$ & 
        $-0.396\pm 0.082$& $0.46^{+0.43}_{-0.49}$\vspace{3 pt} \\
    \end{tabular}
 \caption{Recovered expansion parameters inside balls of redshift $z\leq z^{(\rm cmb)}$. We use the Taylor expansion given in \eqref{eq:agnostic:taylor_expansion_cmb} and the identities \eqref{e:H0} to \eqref{e:j0}. \\
 ~~ \label{tab:agnostic_taylor_z_cmb}}
\end{table}
\end{minipage}
\hspace{0.3 cm}\begin{minipage}{0.485\linewidth}
\begin{figure}[H]
\centering
  \includegraphics[scale=0.5]{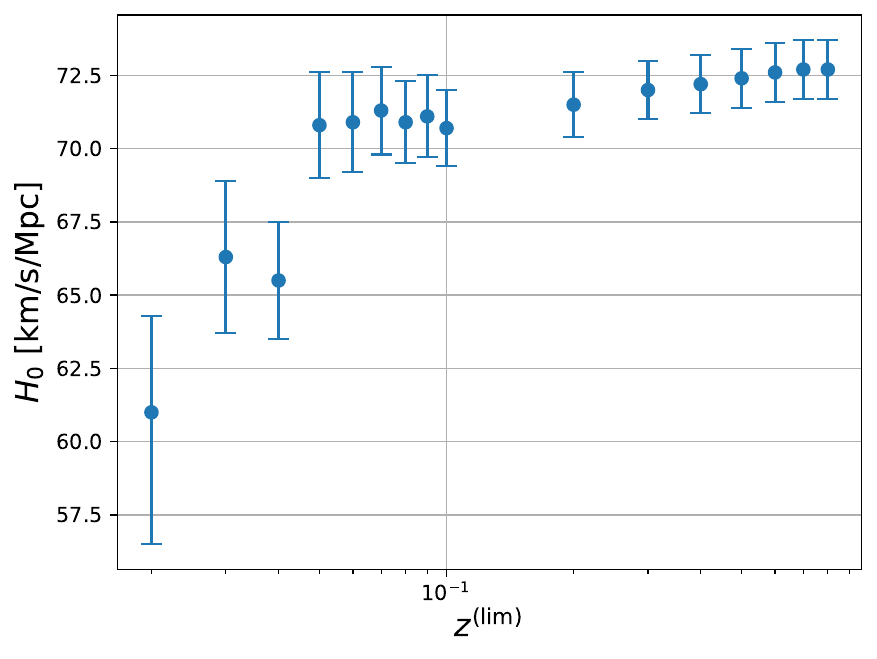}
    \caption{  $H_0$ values for different upper redshift limits using the CMB corrected redshift. At the vertical line, $z^{\rm (lim)}=0.1$, the binning in redshift changes from $0.01$ to $0.1$. Note the jump in $H_0$ at $z^{\rm (lim)}=0.04$.}
  \label{fig:z0_behaviour}
\end{figure}
\end{minipage}

It is also interesting to compare our  values of $q_0=-0.4\pm0.082$ and $j_0=0.46\pm 0.45$ inside a ball of redshift $z= 0.8$ with the value inferred from Planck's best fit flat $\La$CDM model~\cite{Planck:2018vyg} which finds $\Om_m=0.3166\pm0.0084$ or with the Pantheon+ best fit flat $\La$CDM model which obtains $\Om_m=0.334\pm0.018$~:
\bea
\text{Planck :}  \qquad  & q_0 &=~ -0.525\pm 0.013 \\
\text{Pantheon+ :}\qquad &  q_0 &=~ -0.499\pm 0.027   \\
\text{Planck, Pantheon+ :}  \qquad & j_0 &\equiv ~ 1 
 \label{e:j0P+}
\eea
The agnostic value of $q_0$ which we obtain is about $1.5\si$ lower than the Planck best fit, and $1\si$ lower than the Pantheon+ result.  Also $j_0$ is $1.2\si$ lower than the flat $\La$CDM value $j_0=1$. Our errors are larger than the Pantheon+ errors since we have one parameter more. If we compare to the Pantheon+ analysis including curvature which also contains three cosmological parameters, the errors are very similar.

\begin{figure}[!ht]
\begin{subfigure}{.32\textwidth}
  \centering
  \includegraphics[scale=0.3]{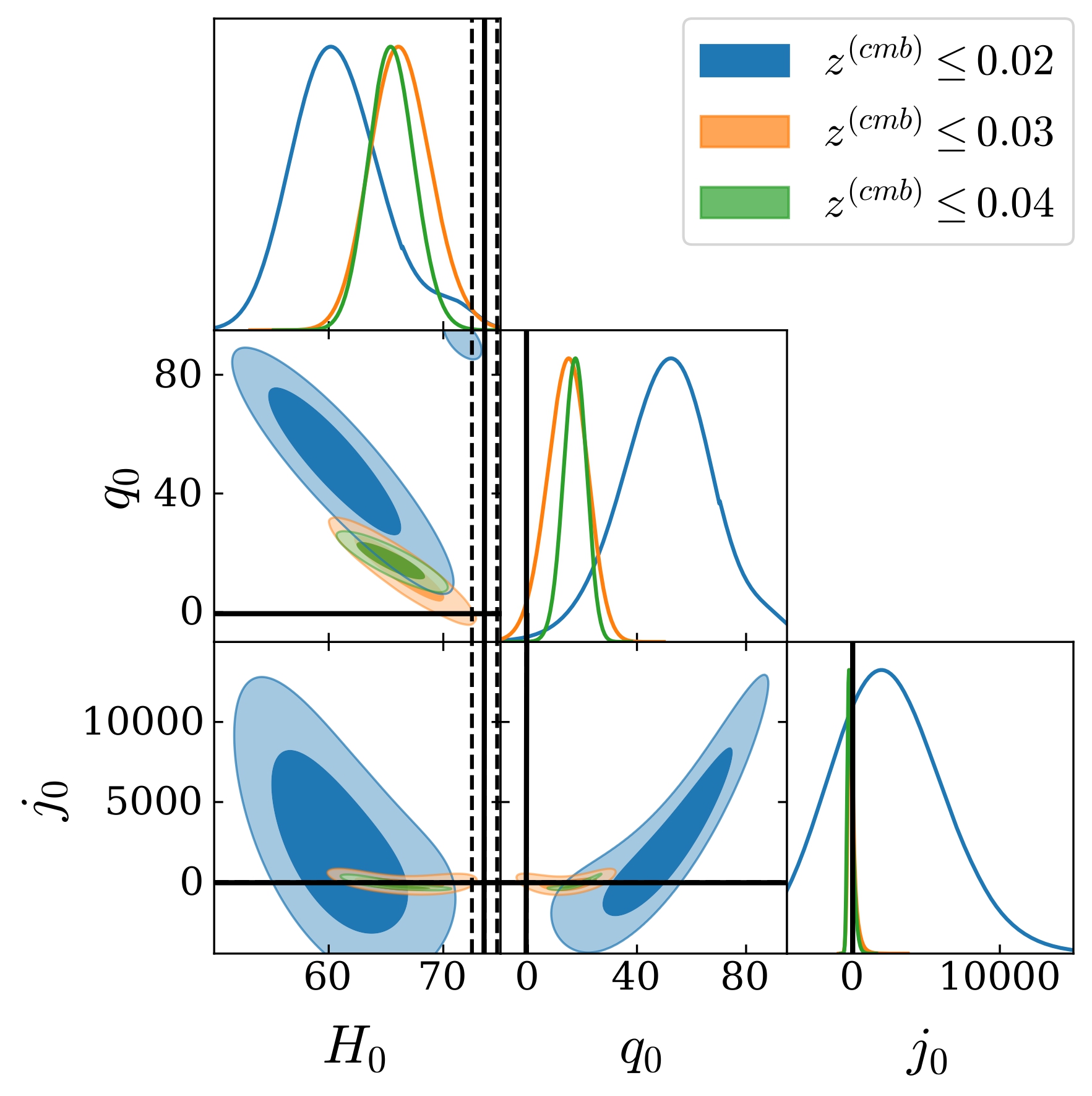}
  \caption{0.02 - 0.04}
  \label{fig:sub-comparison-0.02-0.04}
\end{subfigure}
\begin{subfigure}{.32\textwidth}
  \centering
  \includegraphics[scale=0.3]{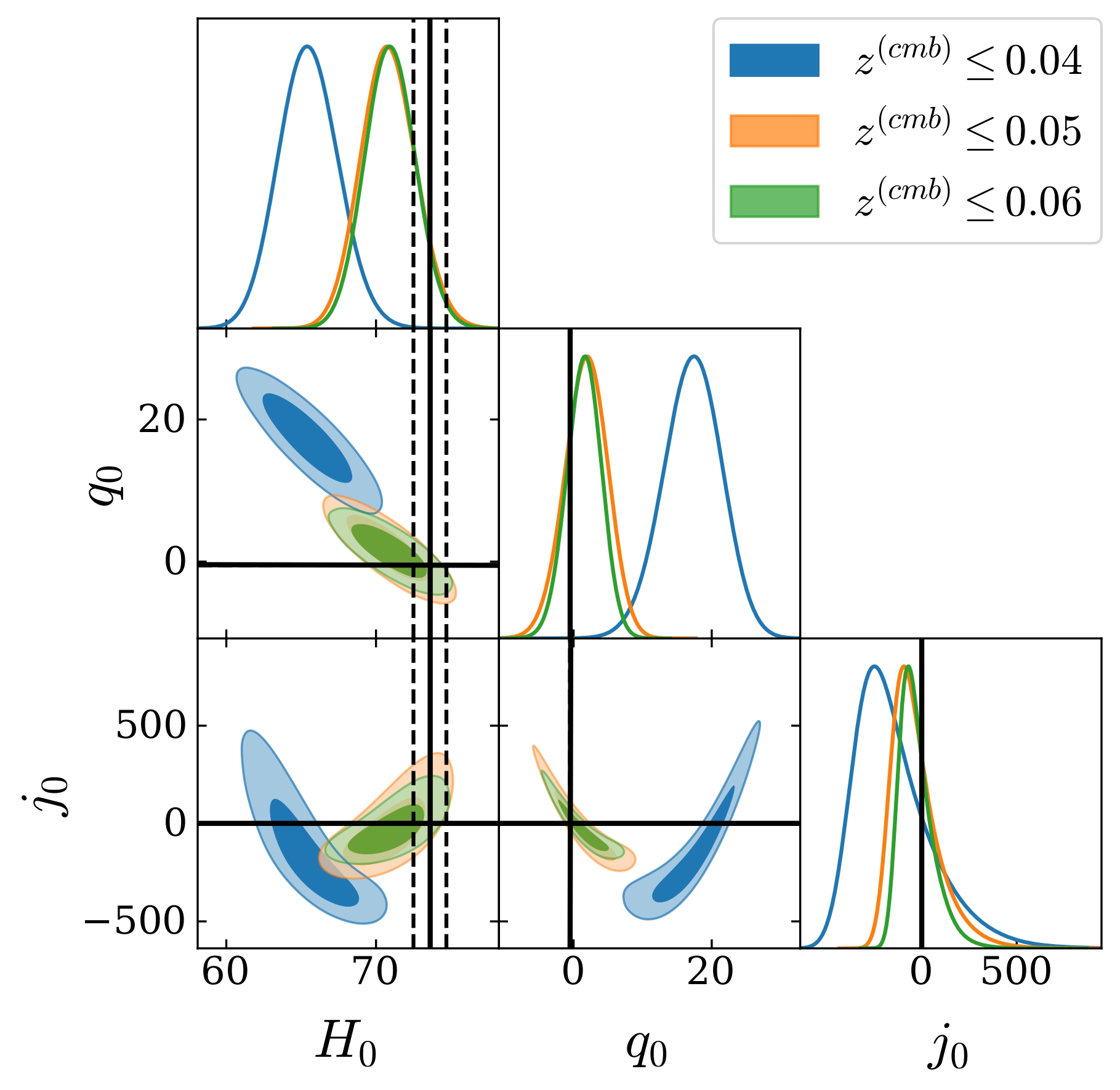} 
  \caption{0.04 - 0.06}
  \label{fig:sub-comparison-0.04-0.06}
\end{subfigure}
\begin{subfigure}{.32\textwidth}
  \centering
  \includegraphics[scale=0.3]{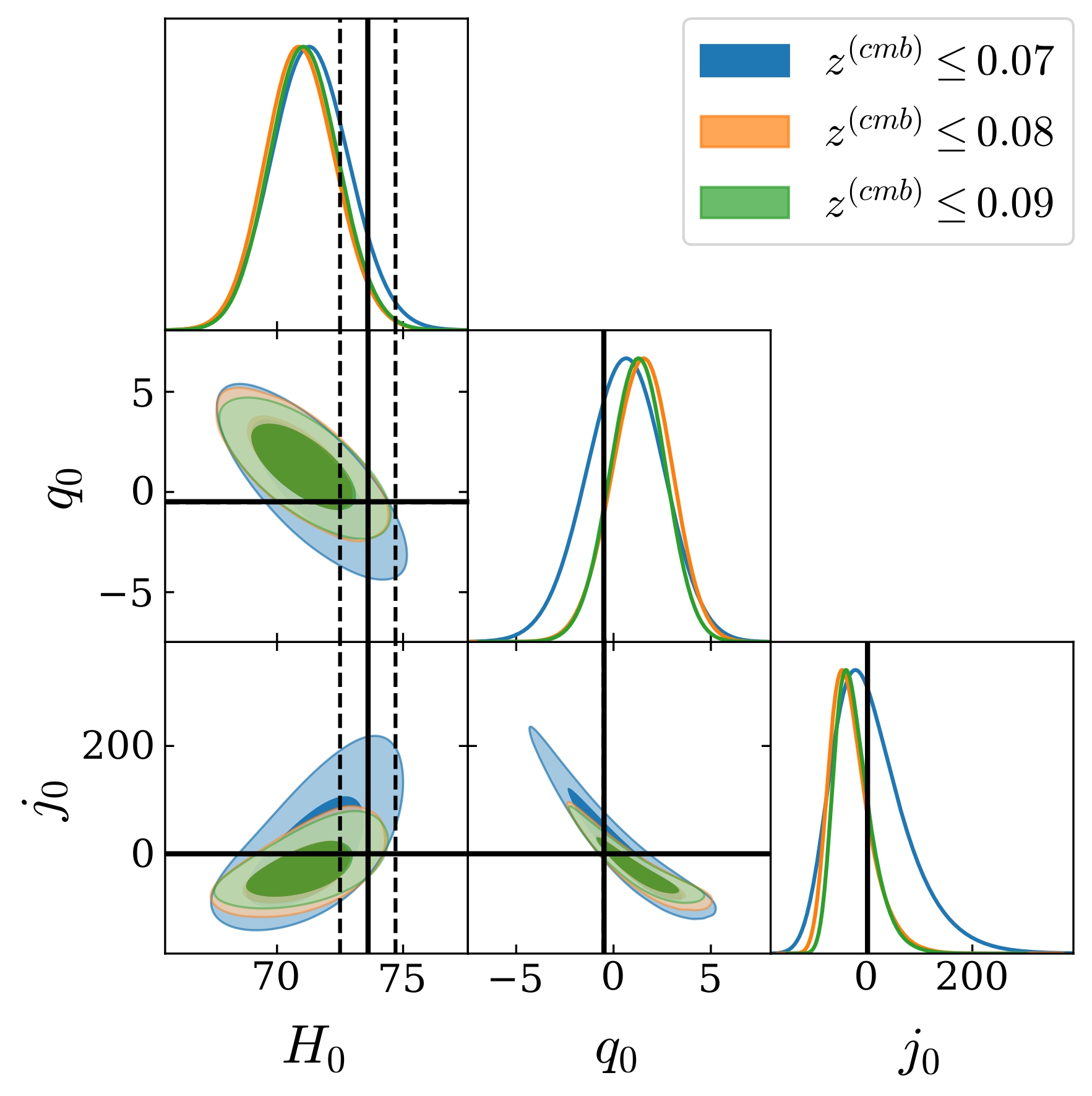}  
  \caption{0.07 - 0.09}
  \label{fig:sub-comparison-0.07-0.09}
\end{subfigure}
\vspace{0.6cm}\\ 
\begin{subfigure}{.32\textwidth}
  \centering
  \includegraphics[scale=0.3]{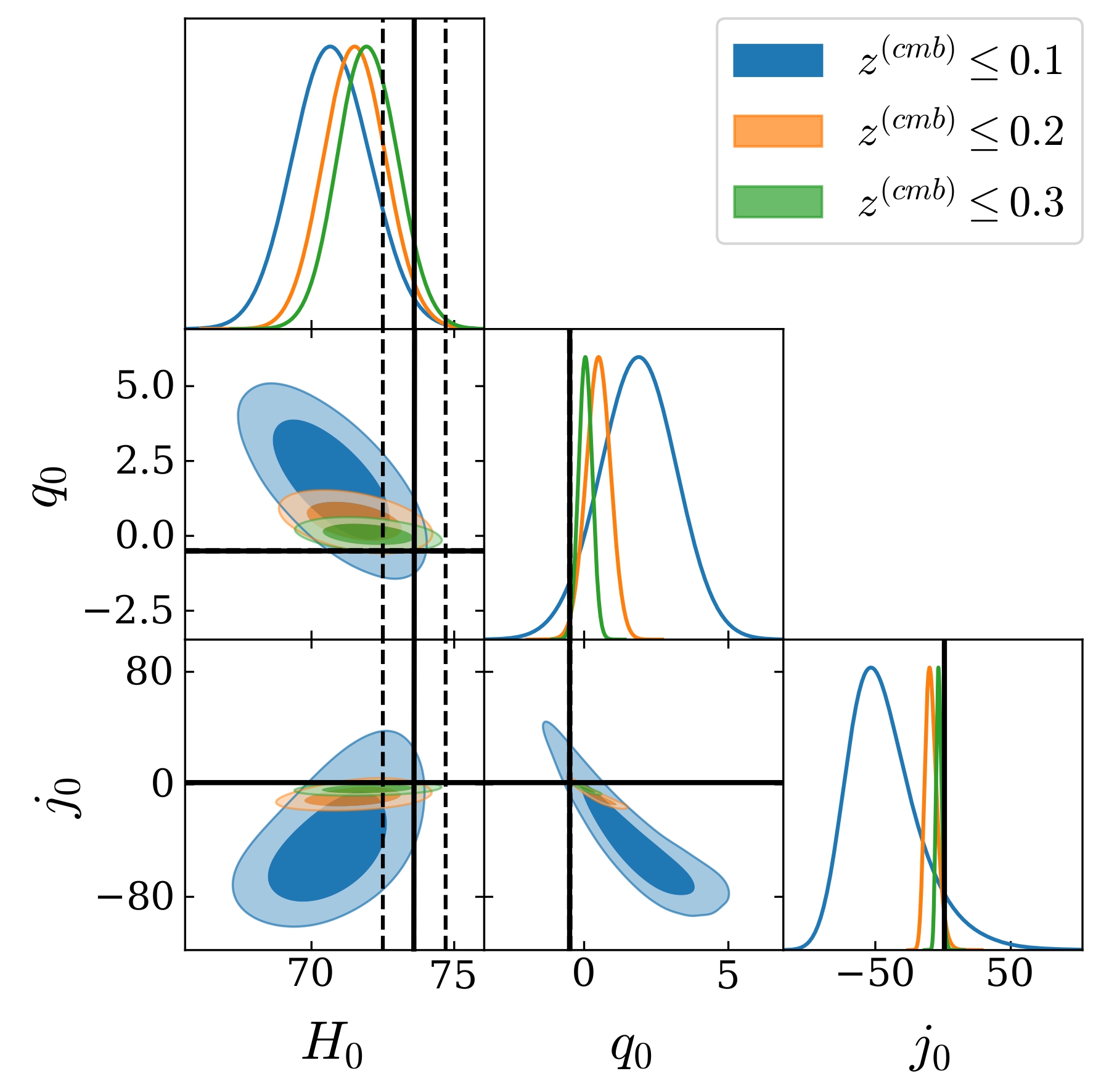}  
  \caption{0.1 - 0.3}
\label{fig:sub-comparison-0.1-0.3}
\end{subfigure}
\begin{subfigure}{.32\textwidth}
  \centering
  \includegraphics[scale=0.3]{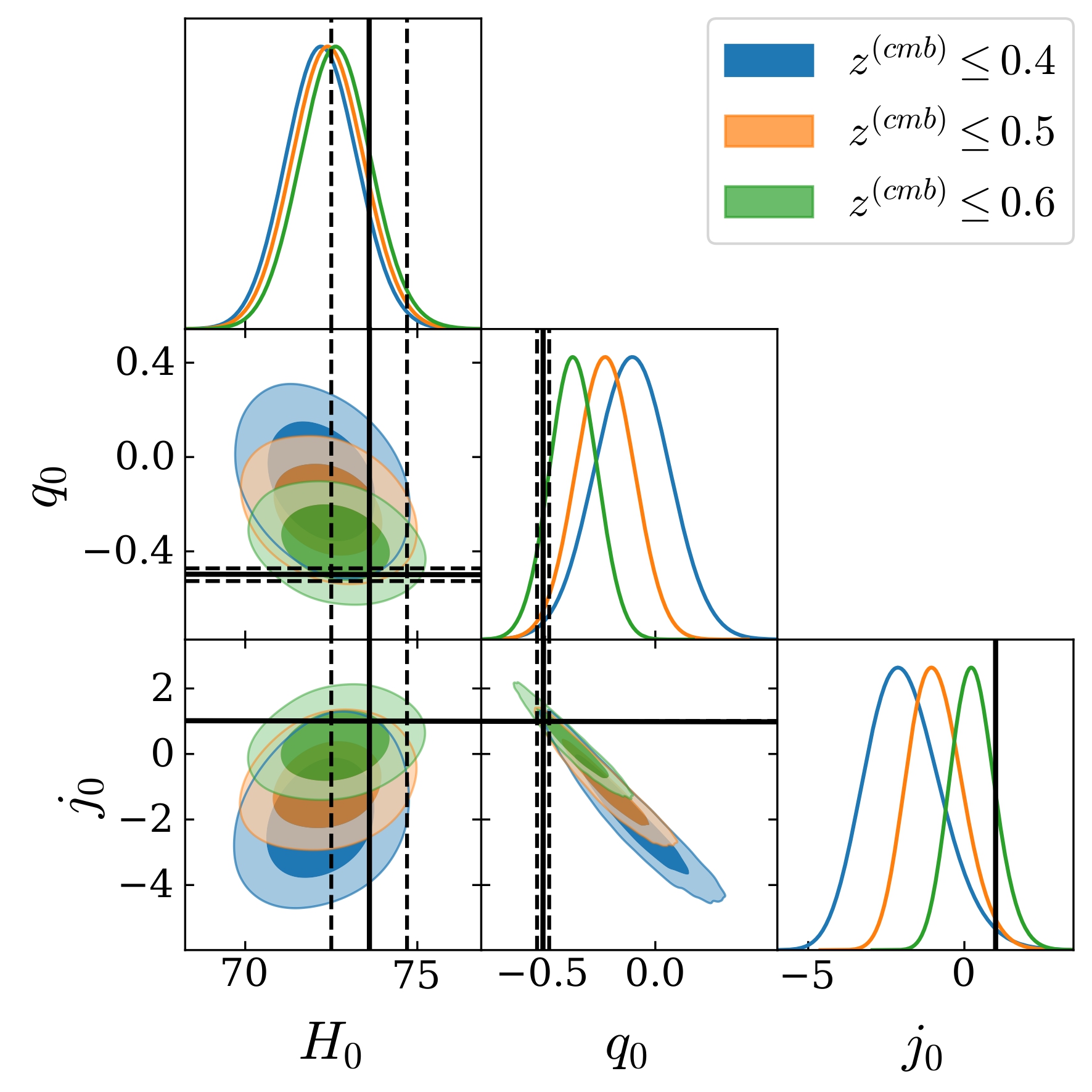} 
  \caption{0.4 - 0.6}
\label{fig:sub-comparison-0.4-0.6}
\end{subfigure}
\begin{subfigure}{.32\textwidth}
  \centering
  \includegraphics[scale=0.3]{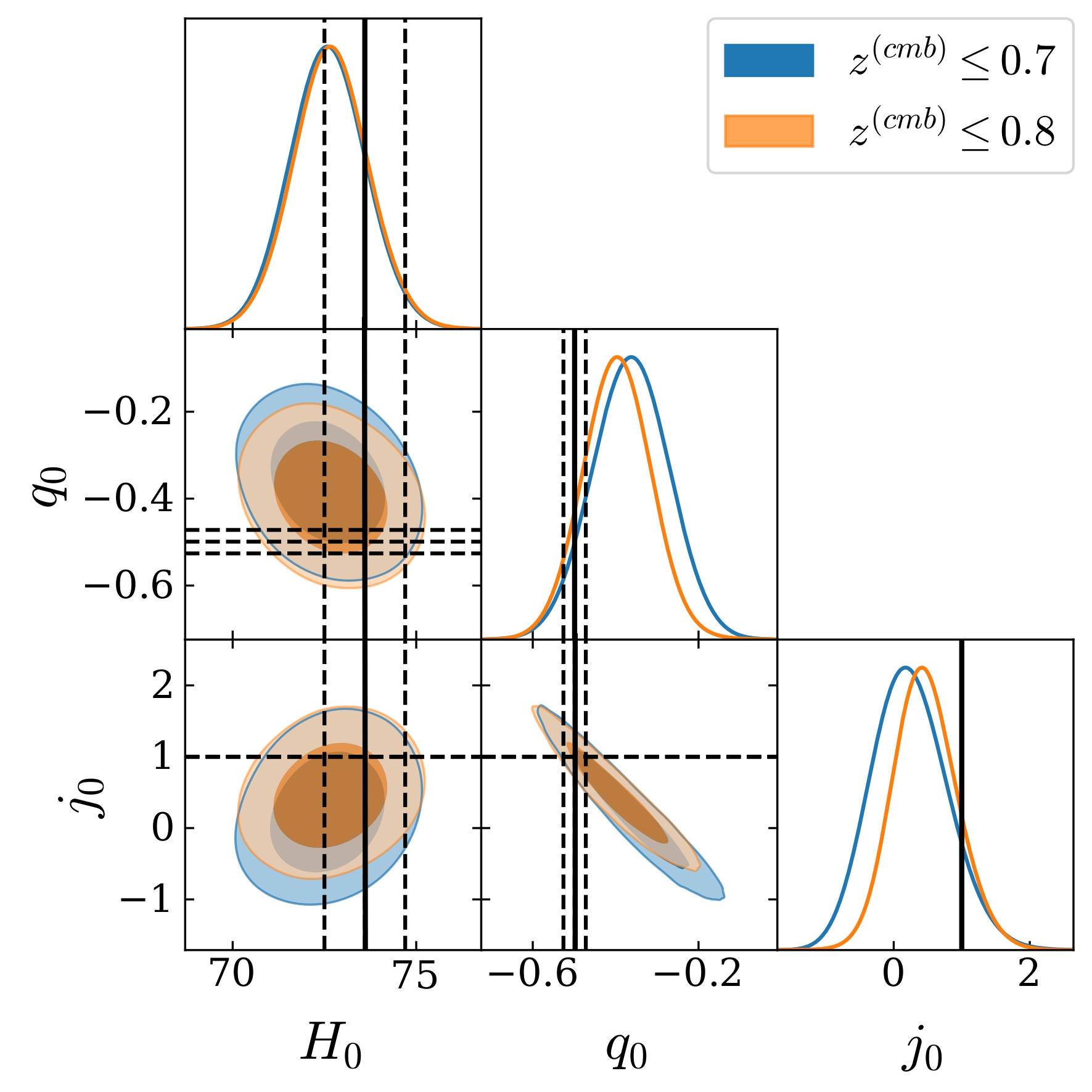}  
  \caption{0.7 - 0.8}
  \label{fig:sub-comparison-0.7-0.8}
\end{subfigure}
\caption{\label{f:compar_agnostic_taylor_expansion_cmb} MCMC results from the Pantheon+  data fitted with an agnostic 3rd order Taylor series expansion. We present the inferred expansion coefficients  using CMB corrected redshifts, $z^{(\rm cmb)}$. The  black vertical lines with $1\sigma$ error bands indicated as dashed lines show the Pantheon+ results for a flat $\La$CDM model as obtained in~\cite{Brout:2022vxf}. Specifically we report the fiducial values and the $1\sigma$ contours for $H_0$ and $\Omm$ (reflected in $q_0$) obtained by the Pantheon+ collaboration~\cite{Brout:2022vxf}. There is a significant jump in the $(q_0,j_0)$ contour from $z^{\rm (lim)}=0.04$ to $z^{\rm (lim)}=0.05$ and $z^{\rm (lim)}=0.06$ in panel (b) which is more than $2\si$ discrepant with the higher redshift contours.}
\end{figure}

In Fig.~\ref{f:compar_agnostic_taylor_expansion_cmb} we present the triangle plots of this analysis. It shows that $q_0$ and $j_0$ are strongly degenerate up to the highest redshifts considered. Within balls of low redshift, $z^{\rm (lim)}<0.5$, these parameters are simply not measurable. It is, however, interesting that a significant transition happens from the ball with $z^{\rm (lim)}=0.04$ to higher redshifts especially visible in the $(q_0,j_0)$ contours shown in panel (b) for which the $z^{\rm (lim)}=0.04$ result is more than $2\si$ discrepant with the higher redshift contours and shows an opposite diagonal degeneracy. The reason for this behaviour will become clear when we add a monopole in Section \ref{sec:z_mon_dip} below. Subsequent $(q_0,j_0)$ contours mainly become smaller and smaller and start constraining $q_0$ and $j_0$ in the right ballpark. For $z^{\rm (lim)}>0.04$, the flat $\La$CDM values, inferred in the Pantheon+ analysis~\cite{Brout:2022vxf} and indicated here as dashed lines, are always displaced from the best fit along the degeneracy direction and lie well inside the $2\si$ contours.

\subsection{A redshift comparison}
\begin{figure}[!ht]
\centering
  \includegraphics[scale=0.55]{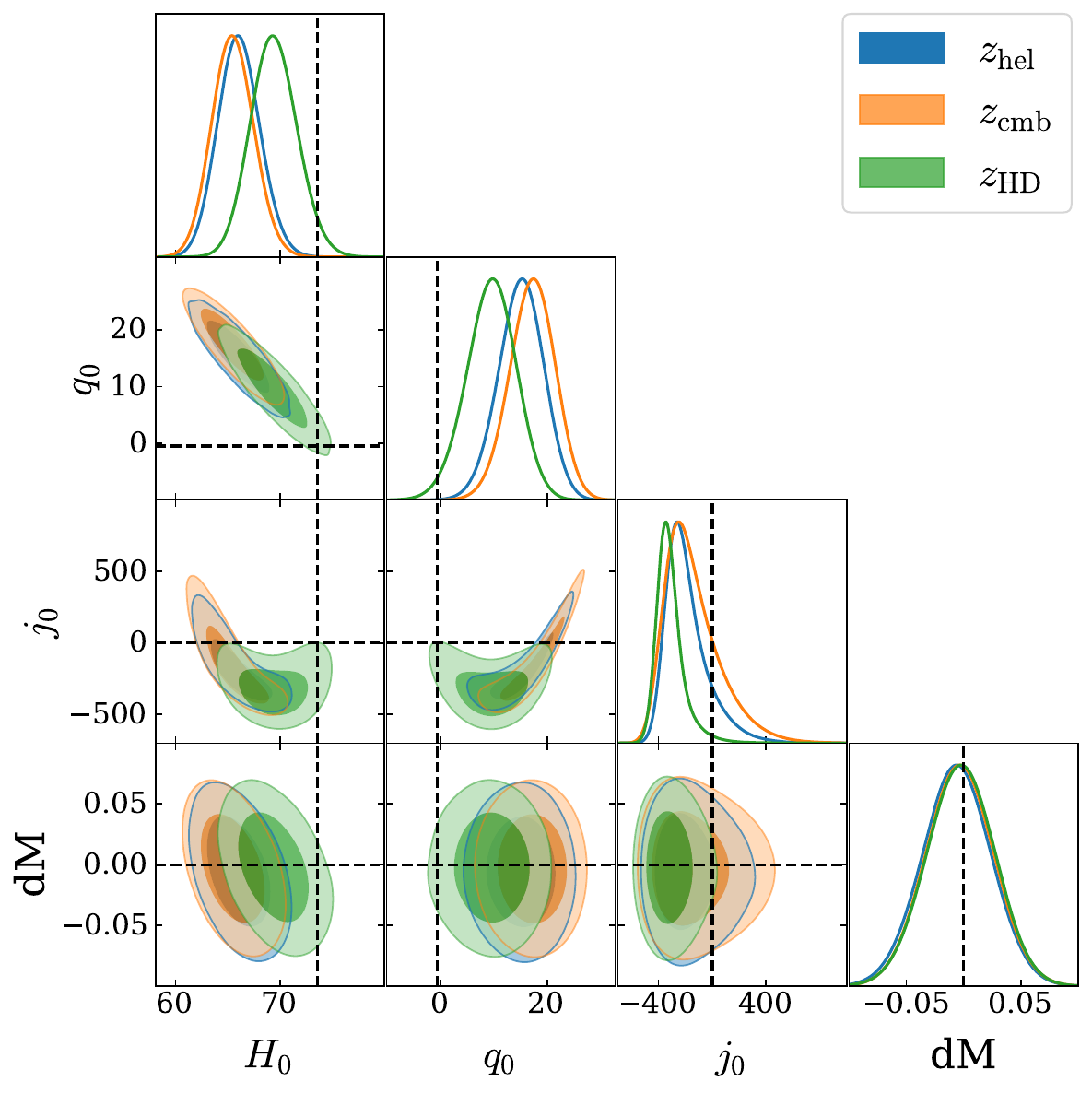}
    \caption{Contour plots for the model independent analysis of Pantheon+ assuming no angular dependence. We are considering all the supernovae with $z<0.04$ using the three different definitions of redshift provided by Pantheon+. The dashed lines show as reference the values obtained by Pantheon+~\cite{Brout:2022vxf}, including $H_0=73.6($km/s/Mpc$)$. }
  \label{fig:tensione_4e2}
\end{figure}

Next we compare the fitted expansion parameters  considering the different redshift definitions that are all provided by the Pantheon+ data release. We choose heliocentric redshifts, $z_{\rm hel}$, CMB corrected redshifts, $z_{\rm cmb}$, which include the peculiar velocity of the solar system with respect to the CMB rest frame, and finally redshifts denoted $z_{\rm HD}$ (Hubble diagram redshifts) which are also corrected for source peculiar velocities modeled by the surrounding galaxy distribution and described in detail in \cite{Brout:2022vxf,Carr_redshift_pantheon+}.  These are the redshifts used in the original Pantheon+ analysis~\cite{Brout:2022vxf}. As we see in Fig.~\ref{fig:tensione_4e2}, where we consider all the SNe with redshift below 0.04,  the contours of all expansion parameters are significantly affected by this choice. Especially the value of $H_0$, while similar for both heliocentric and CMB redshifts, is significantly larger when considering $z_{\rm HD}$. Of course $q_0$ and $j_0$ are badly constrained for this low redshift cut, but even for the $z_{\rm HD}$ redshifts used in the Pantheon+ analysis~\cite{Brout:2022vxf} they are compatible with the $(q_0,j_0)$ values inferred from the full Pantheon+ only just at $2\si$. Nevertheless, the fact that $j_0$ is somewhat better constrained with the $z_{\rm HD}$ analysis indicates that part of the spread of $j_0$ is due to peculiar velocities of the sources. Note, however that for HD redshifts the $\La$CDM value $j_0=1$ is excluded at more than $2\si$. Again, the dashed lines are the values of the original Pantheon+ analysis for a flat $\La$CDM model. Note also that $dM$ is not affected by peculiar velocities since the same velocity is considered for both, the Cepheid and the SNIa in the same galaxy. 

We have also studied the redshift dependence of the effect of peculiar velocities at higher redshifts and we found that their effect becomes negligible for $z^{\rm (lim)}\gsim0.1$.

\section{Including velocities in the redshift definition}\label{s:res}

\subsection{Redshift dipole} \label{subsec:only_dipole}
In order to remain agnostic, instead of imposing the dipole inferred from our velocity with respect to the CMB rest frame, we use the heliocentric redshift z and correct it with an arbitrary dipole, $\bn \cdot \bv_0$. Following the dipole correction described in~\cite{Carr_redshift_pantheon+, Sorrenti:2024} we set
\be \label{theory:z_dipole}
z^{\text{(cor)}}(z,\bn)+1 =\frac{1+z}{1+\de z}\,,
\ee
where
\be \label{theory:z_sun}
1+\de z=\sqrt{\frac{1 -\bn \cdot \bv_0}{1 +\bn \cdot \bv_0}} \,.
\ee
 The position of the supernova in the sky is  $\bn$, while $\bv_0$ is a free velocity vector which we  fit to the dipole of the data. This velocity will be the difference between the velocity of the sun with respect to the CMB rest frame and a possible `bulk velocity' common to all supernovae inside a given redshift ball. As already found previously~\cite{Sorrenti_2023}, at low redshift this bulk velocity can be considerable. We expect to obtain similar results as in~\cite{Sorrenti_2023}, 
where we found $|\bv_0|=318$km/s and (ra, dec)=(140\textdegree, 42\textdegree).

\begin{table}[!ht]
\small
\centering
    \setlength\tabcolsep{9.8pt} 
    \begin{tabular}{ccccccc}
        \toprule
		$z^{\rm (lim)}$ &$H_0$& $q_0$ & $j_0$& $|\bv_0|$ & ra& dec\\ 
	&\scriptsize{[km/s/Mpc]}& & &\scriptsize{[km/s]}& \scriptsize{[deg]}&\scriptsize{[deg]}\\
  \midrule
        0.02 & $60.9^{+3.5}_{-4.3}$ & $43^{+20}_{-10}$ & $2149^{+1000}_{-3000}$ & $252\pm 50$& $164^{+16}_{-19}$& $47\pm 10$\vspace{6 pt} \\        
        0.03 & $66.7\pm 2.7$ & $15.4^{+7.9}_{-7.1}$ & $-115^{+140}_{-300}$ & $294\pm 40$& $142.8^{+9.4}_{-11}$& $45^{+9}_{-8}$\vspace{6 pt} \\
        0.04 & $67.2\pm 2.1$ & $13.9\pm 4.3$ & $-256^{+77}_{-160}$ & $284\pm 40$& $143.7^{+8.8}_{-10}$& $43\pm 8$\vspace{6 pt} \\
        0.05 & $71.9\pm 1.9$ & $0.9\pm 3.2$ & $-17^{+88}_{-170}$& $302\pm 40$ & $139.4^{+7.9}_{-8.8}$&$41^{+8}_{-7}$\vspace{6 pt} \\
        0.06& $72.1\pm 1.8$ & $0.5\pm 2.6$ & $-9^{+71}_{-120}$ & $301\pm 40$& $139.5^{+7.6}_{-8.5}$& $41^{+8}_{-7}$\vspace{6 pt} \\
        0.07& $72.4\pm 1.6$ & $-0.3\pm 2.0$ & $21^{+60}_{-90}$ & $302\pm 40$& $139.3^{+7.6}_{-8.6}$& $40^{+8}_{-7}$ \vspace{6 pt} \\
        0.08 & $72.0\pm 1.4$ & $0.7\pm 1.6$ & $-23^{+32}_{-54}$ & $306\pm 40$& $138.7^{+7.4}_{-8.3}$& $41\pm 7$ \vspace{6 pt} \\
        0.09& $72.3\pm 1.4$ & $0.2\pm 1.4$ & $-4^{+31}_{-51}$ & $308\pm 40$& $138.8^{+7.5}_{-8.4}$& $41\pm 7$\vspace{6 pt} \\
        0.10& $71.9\pm 1.4$ & $0.9\pm 1.4$ & $-28^{+25}_{-41}$ & $307\pm 40$& $139.0^{+7.3}_{-8.3}$& $41^{+8}_{-7}$\vspace{6 pt} \\
        0.20 & $72.2\pm 1.1$ & $0.17\pm 0.44$ & $-5.5^{+4.3}_{-6.3}$ & $307\pm 40$& $139.4\pm 7.9$& $40\pm 7$\vspace{6 pt} \\
        0.30& $72.5\pm 1.1$ & $-0.05\pm 0.24$ & $-2.6^{+1.9}_{-2.4}$ & $305\pm 40$& $139.6\pm 7.8$& $40\pm 7$\vspace{6 pt} \\
        0.40 & $72.7\pm 1.0$ & $-0.18\pm 0.17$ & $-1.5^{+1.1}_{-1.4}$ & $310\pm 40$& $139.9^{+7.2}_{-8.1}$& $41\pm 7$\vspace{6 pt} \\
        0.50& $72.9\pm 1.0$ & $-0.29\pm 0.13$ & $-0.62^{+0.85}_{-1.0}$ & $313\pm 40$& $140.0\pm 7.7$& $41^{+7}_{-6}$\vspace{6 pt} \\
        0.60& $73.1\pm 1.0$ & $-0.43\pm 0.11$ & $0.63^{+0.69}_{-0.80}$ & $316\pm 40$& $139.9\pm 7.6$& $41\pm 7$ \vspace{6 pt} \\
        0.70 & $73.1\pm 1.0$ & $-0.421\pm 0.095$ & $0.52^{+0.55}_{-0.65}$ & $318\pm 40$& $139.7\pm 7.6$& $41^{+7}_{-6}$\vspace{6 pt} \\
        0.80 & $73.1\pm 1.0$ & $-0.448\pm 0.082$ & $0.71^{+0.46}_{-0.52}$ & $316\pm 40$& $139.8\pm 7.7$& $41^{+7}_{-6}$
    \end{tabular}
     \caption{The redshift dipole for SNIa  inside a ball of maximum redshift $z^{\rm (lim)}$. In addition to the best fit expansion parameters we also indicate the velocity amplitude and direction. Even though we do not report it in the table, we also fit for $dM$ which is always very close to zero with an error of $0.03$. \label{tab:params_dipole}}
\end{table}

\begin{figure}[!ht]
\centering
  \includegraphics[scale=0.45]{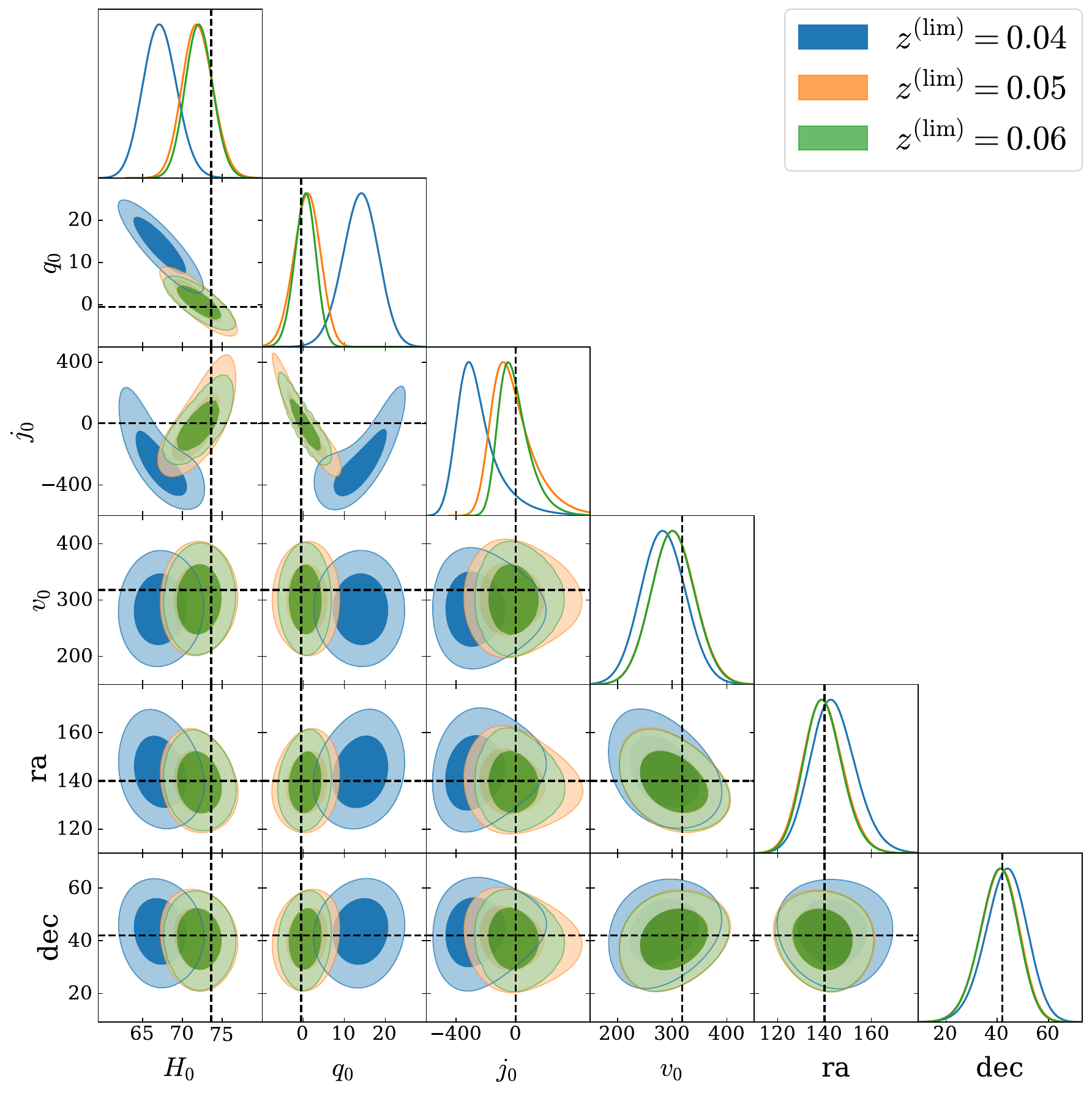}
    \caption{The triangle plot for the expansion parameters and the dipole obtained in an analysis for three redshift balls with limiting redshifts $0.04$, $0.05$ and $0.06$.}
  \label{fig:agnostic_redshift_dipole_hel_from_4e2_to_6e2}
\end{figure}

We report the constraints obtained with our agnostic analysis in Table~\ref{tab:params_dipole} and show them for  limiting redshifts in the range $0.04 - 0.06$ in Fig.~\ref{fig:agnostic_redshift_dipole_hel_from_4e2_to_6e2}. Even though the magnitude offset $dM$ is always fitted together with the other parameters, we do not show it in the tables and plots for brevity.
Note that, contrary to the previous analyses described in~\cite{Sorrenti_2023} and~\cite{Sorrenti:2024}, we now {\em include} all supernovae {\em below} a given redshift. We find that the resulting velocity $\bv_0$ changes very little above $z^{\rm (lim)}=0.05$. This is in agreement with our previous analysis~\cite{Sorrenti:2024} where we {\em excluded} all supernovae {\em below} a certain redshift $z_{\rm cut}$ and found that the bulk velocity is no longer measurable for $z_{\rm cut}\gsim 0.05$. The fact that $\bv_0$ remains constant above $z^{\rm (lim)}=0.05$ does not mean that all these SNIa have the common bulk velocity $\bv_0$ but it is simply due to the fact that the bulk velocity of higher redshift SNIa is irrelevant and does not affect the best fit $\bv_0$. 

If we split $\bv_0$ into the observer velocity determined by the CMB dipole~\cite{Kogut:1993ag,Planck:2013kqc,Planck:2018nkj,Saha:2021bay},
$$ v_{\rm cmb} = (369\pm 0.9){\rm km/s} \,, \qquad  ({\rm ra,dec}) =(167.942\pm 0.007, -6.944\pm 0.007)\,,
$$
and a bulk velocity, $\bv_{\rm bulk}=\bv_{\rm cmb}-\bv_0$ we obtain: \footnote{Error propagation is computed with the help of the \texttt{uncertainties} python library~\cite{uncertainties}.} 
\be
v_{\rm bulk}= (317\pm 37){\rm km/s} \,, \qquad  ({\rm ra,dec}) =(204\pm 12, -53\pm 7)\,.
\ee
This result is in good agreement with our findings in~\cite{Sorrenti_2023,Sorrenti:2024}.
Note also that, similarly as found in~\cite{Lopes_2024}, the direction of this bulk velocity is very close to the direction of the Shapley supercluster, the largest structure in the local Universe located at redshift $z=0.04$ -- $0.05$ and in direction~\cite{Shapely:2006}
\be
\text{direction Shapley supercluster: } \quad
({\rm ra,dec}) =(200, -30)\,.
\ee

Also when including a dipole due to the peculiar velocity the best fit Hubble parameter from low redshifts remains much lower than the result for a full analysis. Actually, comparing Fig.~\ref{fig:agnostic_redshift_dipole_hel_from_4e2_to_6e2} to panel (b) of Fig.~\ref{f:compar_agnostic_taylor_expansion_cmb} we see that the constraints on $H_0$, $q_0$ and $j_0$
change only slightly by enlarging somewhat the $2\si$ error contours. Also comparing tables \ref{tab:agnostic_taylor_z_cmb} and \ref{tab:params_dipole}
we see that the corresponding values of $H_0$, $q_0$ and $j_0$ agree within $1\si$.

\subsection{Redshift monopole and dipole} \label{sec:z_mon_dip}
\begin{figure}[!ht]
\centering
  \includegraphics[scale=0.4]{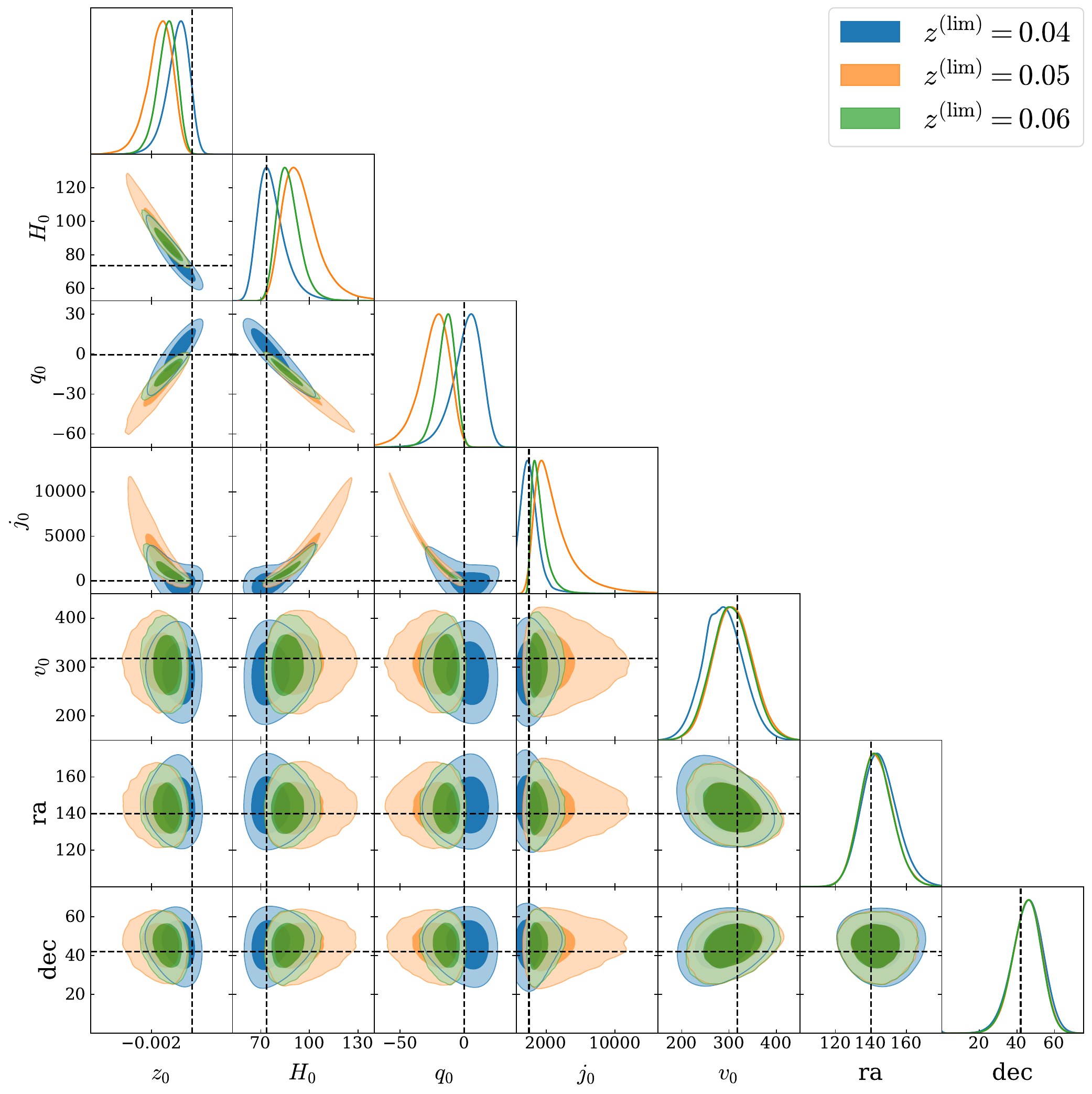}
    \caption{Contour plots for a model assuming both a monopole $z_0$ and a dipole.}
  \label{fig:monopole_dipole_0.04_to_0.06}
\end{figure}

Contrary to a bulk velocity which introduces a dipole on the measured SNIa redshifts, a radial velocity induces a direction independent contribution to the redshift, 
a monopole, which we parameterize as $z_0$. 
The discrepancy we see in $H_0$ and $q_0$ at low redshift can be explained assuming a local overdensity leading to an infall velocity. This results in a negative shift $z_0$ at the level of the redshift, correcting redshift $z^{\text{(cor)}}$ further to $z^{\text{(cor)}}_{1}$ defined by  
\be \label{theory:z_monopole_dipole}
z^{\text{(cor)}}_1(z,\bn)+1 ~=~ \frac{~1+z^{\text{(cor)}}}{1+z_0}\, .
\ee
For this model, we obtain the constraints reported in Table~\ref{tab:params_monopole_dipole} and shown in Fig.~\ref{fig:monopole_dipole_0.04_to_0.06} for the redshift limits $0.04$, $0.05$ and $0.06$.

\begin{table}[!h]
    \setlength\tabcolsep{9pt} 
\small
\centering

    \begin{tabular}{cccccccc}
        \toprule
		 $z^{\rm (lim)}$  &$z_0$&$H_0$& $q_0$ & $j_0$& $v_0$ & ra& dec\\ 
	&\scriptsize{[  $\times 10^{-4}$]}&\scriptsize{[km/s/Mpc]}& & &\scriptsize{[km/s]}& \scriptsize{[deg]}&\scriptsize{[deg]}\\
  \midrule

        0.04 & $-6.7^{+6.0}_{-5.2}$& $78.5^{+6.1}_{-12}$ & $3\pm18$ & $89^{+437}_{-503}$ & $288\pm 40$& $143^{+12}_{-8.5}$& $45^{+9}_{-7}$\vspace{6 pt} \\
        0.05 & $-16.2^{+7.4}_{-5.1}$& $95.3^{+7.8}_{-13}$ & $-24.7^{+14}_{-8.9}$ & $3183^{+1000}_{-3000}$ & $308\pm 40$& $143.3^{+8.6}_{-9.8}$& $45^{+8}_{-7}$ \vspace{6 pt} \\
        0.06 & $-12.4^{+5.4}_{-4.2}$& $87.3^{+5.5}_{-8.1}$ & $-14.8^{+8.1}_{-5.9}$ & $1264^{+500}_{-1000}$ & $305\pm 40$& $143.2^{+8.5}_{-9.7}$& $45^{+8}_{-7}$\vspace{6 pt} \\
        0.07 & $-10.2^{+4.3}_{-3.6}$& $83.4^{+4.3}_{-5.6}$ & $-10.3^{+5.2}_{-4.2}$ & $667^{+300}_{-500}$ & $309\pm 40$& $142.8^{+8.3}_{-9.6}$& $45^{+8}_{-7}$\vspace{6 pt} \\
        0.08 & $-7.3^{+3.6}_{-3.0}$& $78.8^{+3.2}_{-4.0}$ & $-5.0^{+3.4}_{-2.9}$ & $207^{+100}_{-200}$ & $314\pm 40$& $141.6^{+8.1}_{-9.3}$& $45^{+8}_{-7}$ \vspace{6 pt} \\
        0.09 & $-7.4^{+3.4}_{-3.0}$& $79.0^{+3.1}_{-3.7}$ & $-5.2^{+3.0}_{-2.6}$ & $213^{+100}_{-200}$ & $313\pm 40$& $142.1^{+8.2}_{-9.2}$& $45^{+8}_{-7}$ \vspace{6 pt} \\
        0.10 & $-6.1^{+3.2}_{-2.8}$& $77.3^{+2.8}_{-3.4}$ & $-3.3^{+2.8}_{-2.4}$ & $111^{+70}_{-100}$ & $314\pm 40$& $141.7^{+8.0}_{-9.3}$& $45^{+8}_{-7}$ \vspace{6 pt} \\
        0.20 & $-3.3\pm 2.0$& $74.0\pm 1.6$ & $-0.56\pm 0.63$ & $3.1^{+7.1}_{-11}$ & $308\pm 40$& $142.2^{+7.8}_{-8.9}$& $43^{+8}_{-7}$\vspace{6 pt} \\
        0.30 & $-3.2\pm 1.8$& $73.8\pm 1.3$ & $-0.42\pm 0.32$ & $0.4^{+2.6}_{-3.7}$ & $304\pm 40$& $142.9^{+8.2}_{-9.2}$& $43^{+8}_{-7}$\vspace{6 pt} \\
        0.40 & $-3.7\pm 1.7$& $74.2\pm 1.3$ & $-0.49\pm 0.22$ & $0.5^{+1.6}_{-2.2}$ & $306\pm 40$& $143.9^{+8.1}_{-9.1}$& $43^{+8}_{-7}$ \vspace{6 pt}\\
        0.50 & $-3.8\pm 1.7$& $74.4\pm 1.2$ & $-0.54\pm 0.17$ & $0.96^{+1.2}_{-1.5}$ & $310\pm 40$& $143.7^{+8.1}_{-9.1}$& $44^{+8}_{-7}$\vspace{6 pt}\\
        0.60 &$-4.3\pm 1.6$& $74.6\pm 1.2$ & $-0.65\pm 0.14$ & $2.01^{+0.97}_{-1.2}$ & $311\pm 40$& $144.0^{+8.1}_{-9.1}$& $45^{+8}_{-7}$\vspace{6 pt}\\
        0.70 &$-4.0\pm 1.6$& $74.5\pm 1.2$ & $-0.60\pm 0.12$ & $1.53^{+0.75}_{-0.90}$ & $311\pm 40$& $143.7^{+7.9}_{-9.0}$& $44^{+8}_{-7}$\vspace{6 pt}\\
        0.80 &$-4.0\pm 1.5$& $74.5\pm 1.2$ & $-0.60\pm 0.10$ & $1.54^{+0.61}_{-0.74}$ & $310\pm 40$& $143.8^{+8.2}_{-9.3}$& $44^{+8}_{-7}$ 
    \end{tabular}
     \caption{Best fit parameters and their 68\% errors for an agnostic model including a dipole, $\bv_0$ and a monopole, $z_0$ in the redshift. 
     \label{tab:params_monopole_dipole}}
\end{table}
Not surprisingly, there is a significant degeneracy between $H_0$ and $z_0$ at low redshifts. This degeneracy is somewhat reduced at higher redshift due to the different redshift behavior of the terms. Here $z_0$ is assumed constant. 

\begin{figure}[H]
	\centering
	\includegraphics [scale=0.5]{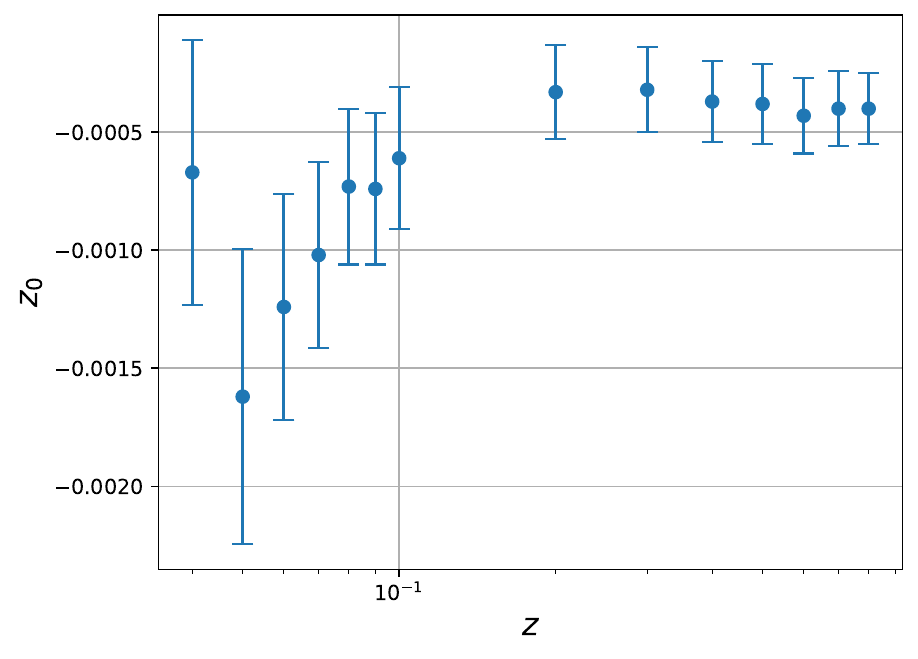}\includegraphics [scale=0.5]{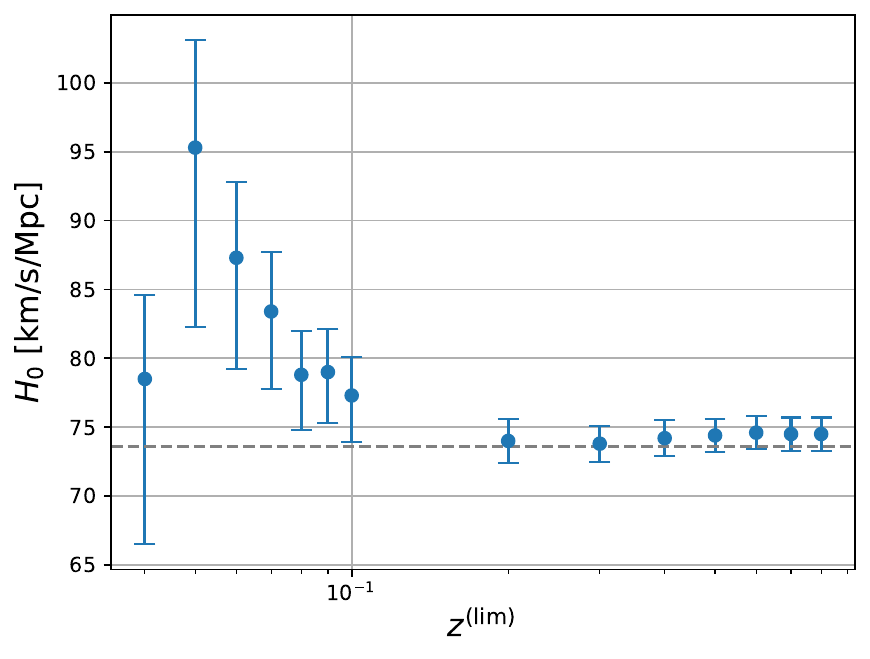}
\caption{$z_0$ (left panel) and $H_0$ (right panel) for the various upper redshift limits using the \textit{dipole+monopole} model described in Sec.~\ref{sec:z_mon_dip} and reported in Table~\ref{tab:params_monopole_dipole}. The dashed horizontal line in the left panel shows the reference values $H_0=73.6$ km/s/Mpc from Pantheon+~\cite{Brout:2022vxf}.\label{f:z0_H0_contours}}
\end{figure}

The behavior of $z_0$ and $H_0$ as functions of the limiting redshift is also illustrated in Fig.~\ref{f:z0_H0_contours}. While low redshift limits, $z^{\rm (lim)}<0.1$ (with a not significant outlier at $z^{\rm (lim)}=0.04$) prefer
a larger negative value of $z_0$ which leads to a larger Hubble constant, for $z^{\rm (lim)}\geq0.2$, the best fit value of $z_0$ settles towards $z_0=-(4\pm1.5)\times 10^{-4}$.  Such a negative redshift (a blueshift) is caused by an infall velocity of $v_r=-(120\pm 45)$km/s. The significance of this infall is somewhat less than 3$\si$. Note also that, once including the monopole $z_0$, the final values for $q_0$ and $j_0$ are in agreement with the Planck values within less than $1\si$.

It is interesting to note that when forcing $z_0=0$, the parameter $d^{(3)}$ is not compatible with zero even at low redshift, where we expect a nearly linear Hubble law. The $d_L(z)$ curve then is significantly bent with $d^{(3)}\sim 10$ for $z^{\rm (lim)}\leq 0.04$. This is also the origin of the jump in $(q_0,j_0)$ from $z^{\rm (lim)}= 0.04$ to $z^{\rm (lim)}= 0.05$ in Fig.\ \ref{f:compar_agnostic_taylor_expansion_cmb} which nearly disappears once we allow for $z_0\neq 0$. Including sufficiently many higher redshift supernovae, the significance of this bend in the $d_L(z)$ curve also disappears.

We have also tested the stability of the result under the addition of a quadrupole or when neglecting the dipole. While the value of $z_0$ becomes somewhat larger when we neglect the dipole (about $6\times 10^{-4}$), the inclusion of a quadrupole does not affect the result.
Of course including a dipole we have a much better fit to the data, $\De\chi^2=77$ than with a monopole only. Including also a quadrupole in the redshift, improves the fit only by  $\De\chi^2=2.3$ which is not significant considering the five additional parameters introduced in the fit.

\subsection{A ``tomographic'' analysis} \label{subsec:tomo}

Next we allow for a redshift dependence of $z_0$. In order not to over-fit the data and loose any constraining power, we allow for four different values of $z_0$:
\be
\begin{array}{ccc}
z_{0,1} & \quad \text{in the redshift interval}\quad & \qquad 0<z\leq 0.04\,,\\
z_{0,2} & \quad \text{in the redshift interval} \quad & 0.04<z\leq 0.1\,,\\
z_{0,3} &\quad  \text{in the redshift interval} \quad & 0.1<z\leq 0.3\,,\\
z_{0,4} &\quad  \text{in the redshift interval}\quad  & 0.3<z\leq 0.8\,.
\end{array}
\ee
The best fit values and their 68\%  confidence intervals for this analysis are reported in Table \ref{tab:agnostic_contours_3d} and the redshift dependence of $z_0$ is also illustrated in Fig.~\ref{f:agnostic_z_0_i}.

\begin{minipage}{0.48\linewidth}
\begin{table}[H]

\centering

\begin{tabular}{ccc}
    \toprule

$10^4z_{0,1}$& $-2.9\pm 1.7       $\\
$10^4z_{0,2}$ & $-0.8\pm 4.4       $\\
$10^4z_{0,3}$& $3\pm 10          $\\
$10^4z_{0,4}$& $-19\pm 22         $\\
$H_0$& $73.8\pm 1.2               $\\
$q_0$ & $-0.51\pm 0.14             $\\
$j_0$ & $1.24^{+0.66}_{-0.87}      $\\
$v_0$ & $302\pm 40                 $\\
ra & $143.2^{+8.1}_{-9.2}       $\\
dec & $43^{+8}_{-7}              $\\
dM & $-0.007\pm 0.029           $\\
\bottomrule

\end{tabular}
\caption{\small Constraints obtained using a model with velocity dipole and a free monopole $z_{0,i}$ in four redshift bins. We are considering all the SNe with $z\leq0.8$.}.\label{tab:agnostic_contours_3d}
\vspace{0.2cm}
\end{table}

\end{minipage}~~
\begin{minipage}{0.48\linewidth}
    \begin{figure}[H]
        \centering
        \includegraphics[scale=0.5]{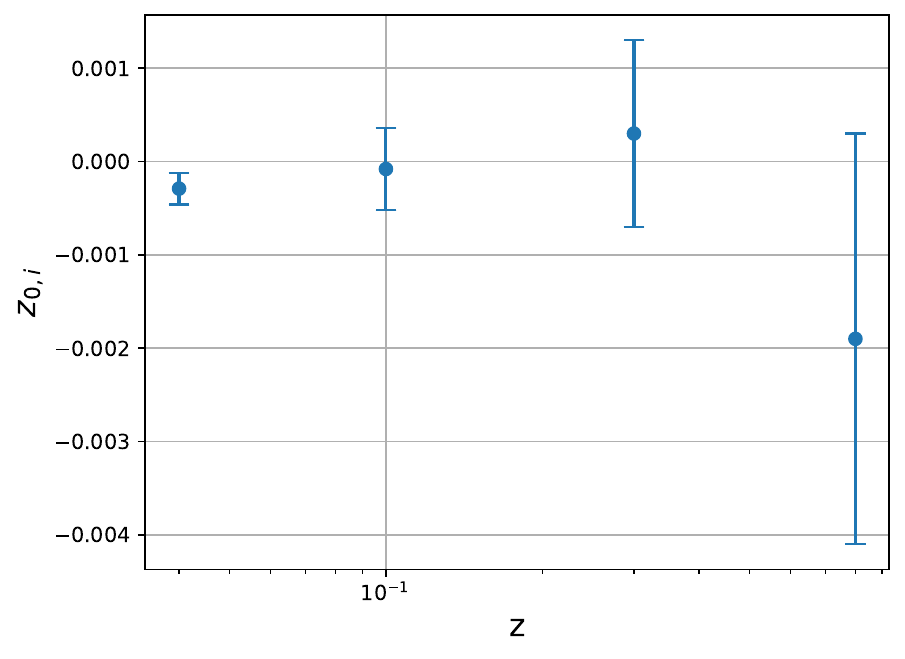}
        \caption{\small The behaviour of the redshift shifts $z_{0,i}$ reported in Table~\ref{tab:agnostic_contours_3d}.  \label{f:agnostic_z_0_i}}
        \vspace{0.17cm}
    \end{figure}
\end{minipage}

Only the lowest bin, $z^{\rm (lim)}=0.04$, requires a non-vanishing blueshift $-z_0$ at more than 1$\si$ significance. This prompts us to consider a model with radial infall only for $z<0.04$.

\begin{figure}[!ht]
\centering
  \includegraphics[scale=0.35]{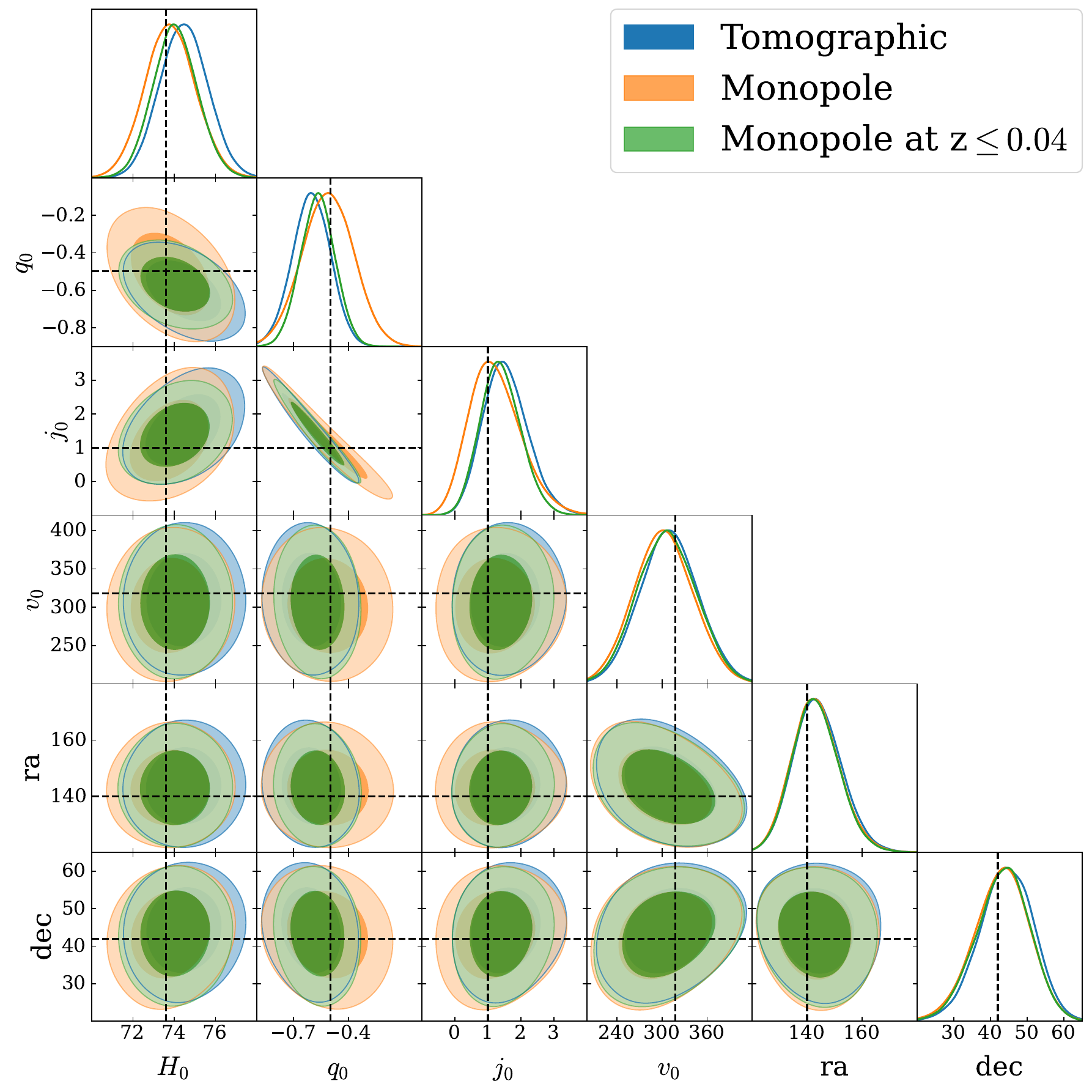}
    \caption{Considering all the SNe with $z\leq0.8$, we perform a comparison between the "tomographic" model,the standard approach fitting "monopole+dipole" and a model fitting a "dipole" with a monopole only for SNe with z$\leq 0.04$. }
  \label{fig:3d_vs_standard_vs_standard_0.04_max}
\end{figure}


\subsection{\texorpdfstring{Redshift dipole and a monopole for SNe at $z \leq 0.04$}{Redshift dipole and a monopole for SNe at z 0.04}} \label{subsec:z_mon_max_04_dip}

Here we test whether a local overdensity leading to a radial infall only for the relatively nearby SNe with $z\leq 0.04$ provides an as good or even better fit than an overall constant blueshift. To study this, we perform  a \textit{dipole+monopole} analysis as in section~\ref{sec:z_mon_dip}, but this time allowing for a non-vanishing monopole $z_0$ only if the supernova $i$ has a redshift $z^i \leq 0.04$, i.e.,
\be \label{e:z0_cases}
z_0^i =
\begin{cases}
z_0,  \quad \hspace{0.3 cm} z^i\leq 0.04,\\
0, \quad \hspace{0.5 cm} \text{otherwise}, \\
\end{cases}
\ee 
where $z_0^i$ is the monopole redshift associated to supernova $i$.

\begin{table}[!ht]
    \setlength\tabcolsep{8pt} 
\small
\centering
 \begin{tabular}{cccccccc}
        \toprule
		 $z^{\rm (lim)}$  &$-z_0$&$H_0$& $-q_0$ & $j_0$& $v_0$ & ra& dec \\ 
	&\scriptsize{[  $\times10^{-4}$]}&\scriptsize{[km/s/Mpc]}& & &\scriptsize{[km/s]}& \scriptsize{[deg]}&\scriptsize{[deg]}\\
  \midrule

        0.04 & $6.7_{-6.0}^{+5.2}$& $78.5^{+6.1}_{-12}$ & $-3\pm18$ & $89^{+437}_{-503}$ & $288\pm 40$& $143^{+12}_{-8.5}$& $45^{+9}_{-7}$\vspace{6 pt} \\
        0.05 & $11.2\pm 3.5$& $84.7^{+4.5}_{-5.5}$ & $10.8_{-6.0}^{+5.1}$ & $662^{+400}_{-700}$ & $307\pm 40$& $143.3^{+8.3}_{-9.6}$& $45^{+8}_{-7}$\vspace{6 pt} \\
        0.06 & $11.1\pm 3.5$& $84.8^{+4.4}_{-5.3}$ & $11.0_{-5.3}^{+4.5}$ & $673^{+300}_{-600}$ & $305\pm 40$& $143.2^{+8.3}_{-9.7}$& $45^{+8}_{-7}$\vspace{6 pt} \\
        0.07 & $11.0\pm3.4$& $85.1^{+4.3}_{-5.2}$ & $11.8_{-5.0}^{+4.2}$ & $763^{+300}_{-500}$ & $305\pm 40$& $143.0^{+8.2}_{-9.5}$& $44^{+8}_{-7}$\vspace{6 pt} \\
        0.08 & $10.6\pm3.5$& $83.6^{+4.1}_{-5.0}$ & $9.7_{-4.5}^{+3.7}$ & $529^{+200}_{-400}$ & $310\pm 40$& $142.4^{+8.2}_{-9.3}$& $45^{+8}_{-7}$\vspace{6 pt} \\
        0.09 & $10.6\pm3.4$& $83.8^{+4.0}_{-4.9}$ & $9.9_{-4.3}^{+3.6}$ & $546^{+200}_{-400}$ & $312\pm 40$& $142.7^{+8.2}_{-9.3}$& $45^{+8}_{-7}$ \vspace{6 pt} \\
        0.10 & $9.5^{+3.5}_{-3.2}$& $81.9^{+3.7}_{-4.6}$ & $7.8_{-4.0}^{+3.3}$ & $366^{+200}_{-300}$ & $312\pm 40$& $143.0^{+8.1}_{-9.2}$& $45^{+8}_{-7}$\vspace{6 pt} \\
        0.20 & $3.0\pm1.7$& $74.0\pm 1.5$ & $0.76\pm 0.70$ & $6.9^{+8.4}_{-13}$ & $307\pm 40$& $142.2^{+7.8}_{-8.9}$& $43^{+8}_{-7}$\vspace{6 pt} \\
        0.30 & $2.7\pm1.4$& $73.7\pm 1.2$ & $0.44\pm 0.32$ & $0.8^{+2.8}_{-3.9}$ & $303\pm 40$& $142.5^{+7.9}_{-9.1}$& $43^{+8}_{-7}$ \vspace{6 pt} \\
        0.40 & $3.0\pm1.3$& $73.9\pm 1.2$ & $0.49\pm 0.22$ & $0.7^{+1.7}_{-2.1}$ & $306\pm 40$& $143.3\pm 8.6$& $43^{+8}_{-7}$ \vspace{6 pt}\\
        0.50 & $3.2\pm1.3$& $74.0\pm 1.1$ & $0.52\pm 0.16$ & $0.9^{+1.2}_{-1.4}$ & $306\pm 40$& $143.7\pm 8.6$& $43^{+8}_{-7}$\vspace{6 pt}\\
        0.60 &$3.2\pm1.3$& $74.2\pm 1.1$ & $0.62\pm 0.13$ & $1.85^{+0.92}_{-1.1}$ & $308\pm 40$& $143.7^{+8.0}_{-9.0}$& $44^{+8}_{-7}$ \vspace{6 pt}\\
        0.70 &$3.3\pm1.2$& $74.1\pm 1.1$ & $0.57\pm 0.11$ & $1.37^{+0.69}_{-0.82}$ & $306\pm 40$& $143.8^{+7.9}_{-9.2}$& $43^{+8}_{-7}$\vspace{6 pt}\\
        0.80 &$3.2\pm1.2$& $74.0\pm 1.1$ & $0.569\pm 0.094$ & $1.39^{+0.57}_{-0.65}$ & $307\pm 40$& $143.2^{+8.0}_{-8.9}$& $43^{+8}_{-7}$
    \end{tabular}
     \caption{Contours obtained performing a \textit{dipole+monopole} analysis as described in section~\ref{subsec:z_mon_max_04_dip},  where we fit the monopole considering only the SNe with redshift below 0.04. \label{tab:params_monopole_max_04_dipole}}
\end{table}

We obtain the best fits and errors reported in Table~\ref{tab:params_monopole_max_04_dipole}.
In Fig.~\ref{fig:3d_vs_standard_vs_standard_0.04_max} we compare the contours for the expansion parameters and the dipole of this fit with the tomographic analysis and with the one allowing for a constant redshift monopole $z_0$ in all supenovae when considering all supernovae with $z\leq 0.8$ in the Pantheon+ data. The new fit with $z_0\neq 0$ only for $z\leq 0.04$  outperforms the constant $z_0$ and is not worse than the tomographic analysis which has three more parameters. This is indicated by the better constraints on the parameters considered in  Fig.~\ref{fig:3d_vs_standard_vs_standard_0.04_max}.  A more precise measure of the goodness of fit is the $\chi^2$ difference. The value of $\chi^2$ for this model is lower 
than the constant monopole for all limiting redshifts as shown in Table~\ref{tab:delta_chi_mono_momo_max},
where we report 
the $\chi^2$ difference of the model introducing a constant $z_0$ for all supernovae as discussed
 in Section~\ref{sec:z_mon_dip} and this model. While $\De\chi^2$ is positive for all limiting redshifts $z$, the improvement is significant only for $z^{\rm (lim)}\leq0.1$. For higher redshift limits, where the Hubble parameter is well determined and has settled to its final value, see Fig~\ref{f:H0_values_mono}, the resulting $\De\chi^2$  is always less than one, indicating that the data no longer distinguishes between the two models. From Table~\ref{tab:params_monopole_max_04_dipole} we also see that after about $z^{\rm (lim)}=0.2$ the redshift $z_0$ settles to its final value of $z_0=-(3.2\pm 1.2)\times 10^{-4}$, which is in the end detected at nearly $3\si$ significance. We also report the $\chi^2$ difference of this model with the best fit model without a monopole. We find that introducing a monopole $z_0\neq 0$ improves the model significantly with $\De\chi^2 =7.26$. This is a quite significant improvement for the addition of only one  parameter.
\\
\begin{minipage}{0.38\linewidth}

\begin{table}[H]
\small
    \centering
    \setlength\tabcolsep{6pt} 
    \begin{tabular}{c c c}
        & \multicolumn{2}{c}{$\Delta \chi^2$}  
        \vspace{3 pt} \\
        \toprule
        $z^{\rm (lim)}$ & \textit{Dipole+monopole} & \textit{Only} \\
         &\textit{all SNe} & \textit{dipole} \\
        \midrule
        0.05 & $-1.83$& $-11.8$\\
        0.06 & $-3.15$& $-11.8$ \\
        0.07 & $-3.66$& $-11.6$ \\
        0.08 & $-5.37$& $-10.6$  \\
        0.09 & $-5.06$& $-10.9$ \\
        0.1 & $-4.84$& $-8.98$ \\
        0.2 & $-0.42$& $-2.84$ \\
        0.3 & $-0.17$& $-3.34$ \\
        0.4 & $-0.44$& $-4.90$ \\
        0.5 & $-0.51$& $-6.03$ \\
        0.6 & $-0.61$& $-7.71$ \\
        0.7 & $-0.57$& $-7.08$ \\
        0.8 & $-0.44$& $-7.26$ \\        
        \bottomrule
    \end{tabular}
    \caption{\captionsize $\Delta \chi^2$ between the best fit for the \textit{dipole+monopole} model adding a monopole only to SNe with redshift $z\leq 0.04$~(Sec.~\ref{subsec:z_mon_max_04_dip}) and the analysis of Sec.~\ref{sec:z_mon_dip} adding a monopole $z_0$ to all the SNe  (second column) and the analysis of Sec.~\ref{subsec:only_dipole} with only a dipole (third column). }
    \label{tab:delta_chi_mono_momo_max}
\end{table}
\end{minipage}~~
\begin{minipage}{0.58\linewidth}
    \begin{figure}[H]
        \centering
        \includegraphics[scale=0.58]{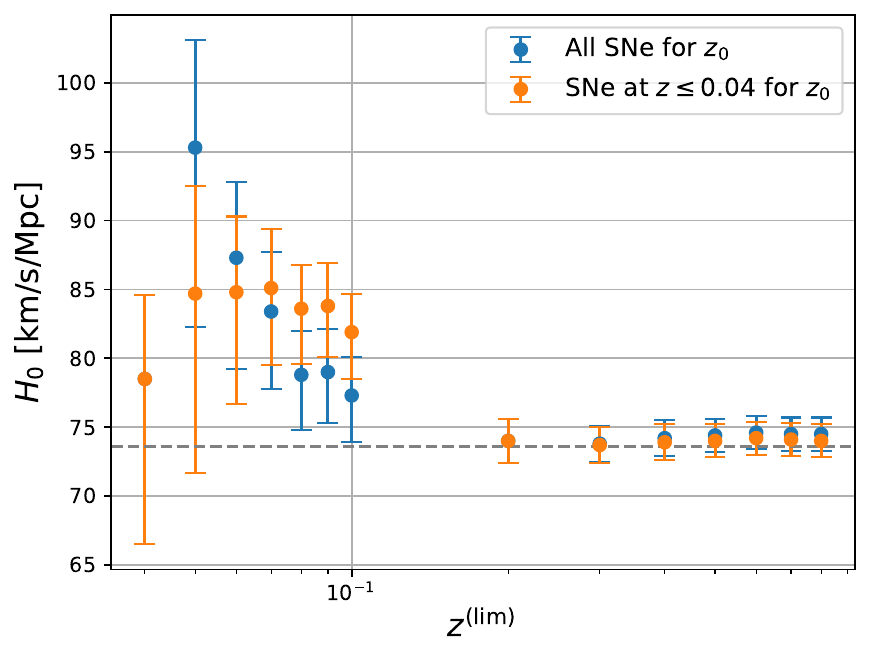}
        \caption{\small $H_0$ values for different upper redshift limits for the \textit{dipole+monopole} analysis performed adding the monopole $z_0$ to all the SNe (sec.~\ref{sec:z_mon_dip}) and adding the monopole only to SNe with redshift $\leq 0.04$~(sec.~\ref{subsec:z_mon_max_04_dip}). The dashed horizontal line shows the reference values $H_0=73.6$ km/s/Mpc from Pantheon+~\cite{Brout:2022vxf}. \label{f:H0_values_mono}}

    \end{figure}
\end{minipage}

\subsection{Comparing to Pantheon+ peculiar velocities}
\begin{figure}[ht]
	\centering
	\includegraphics [scale=0.5]{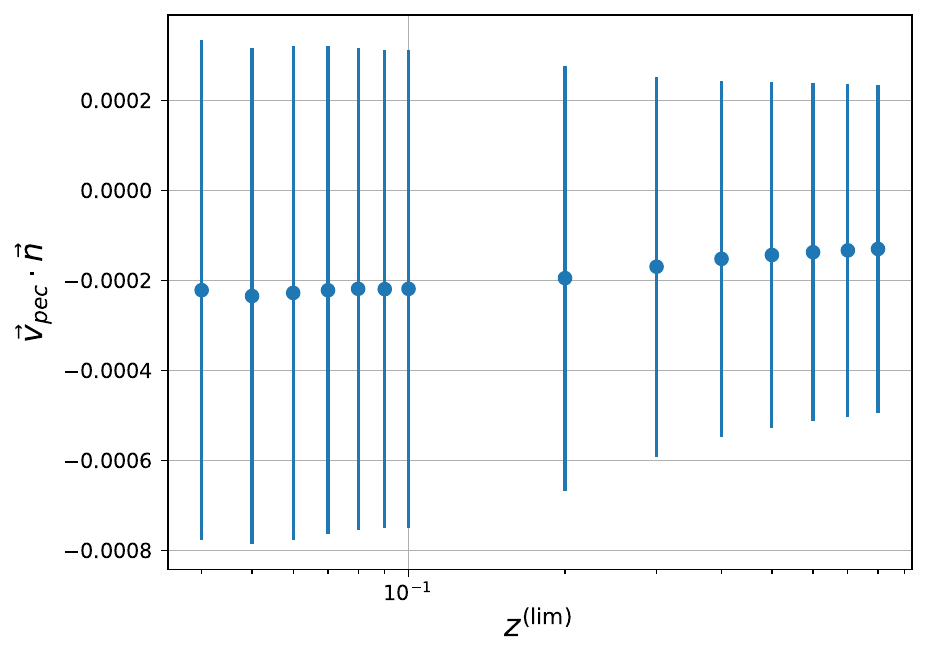}\includegraphics [scale=0.5]{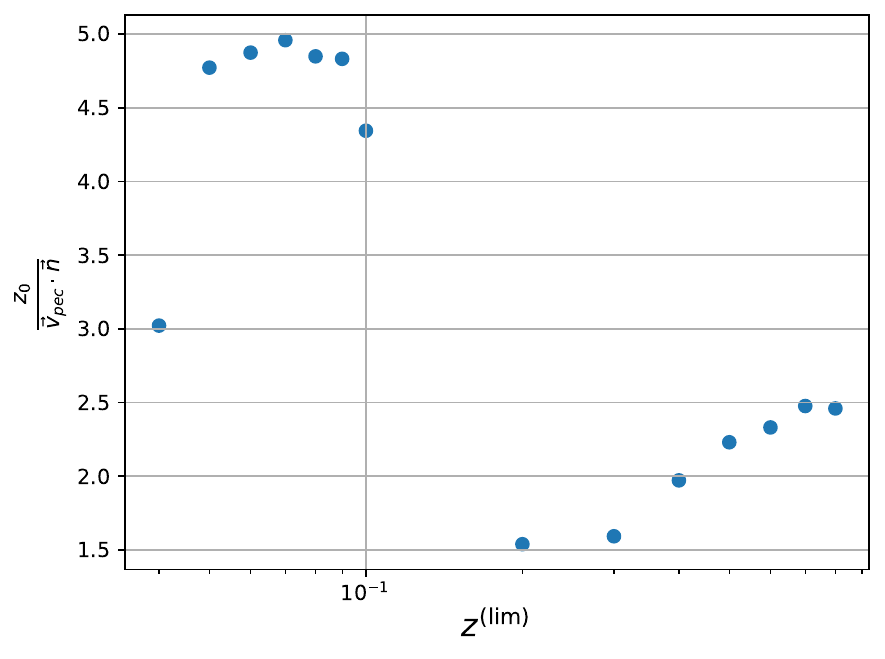}
\caption{Peculiar velocities assumed by Pantheon+ in units where $c=1$ (left panel) and ratio between the same peculiar velocities to the monopole $z_0$  obtained from our agnostic analysis as function of the upper limiting redshift $z^{\rm (lim)}$ (right panel). 
\label{f: vpec_distributions}}
\end{figure}

As we have  mentioned, also the original Pantheon+ analysis allows for peculiar velocities of their supernovae.  Already Fig.~\ref{fig:tensione_4e2} indicates that these velocities must have a significant radial component in order to bring the best fit Hubble constant from 65km/s/Mpc to about 70km/s/Mpc inside the redshift range $0<z\leq 0.04$. To compare these with our analysis we average the radial peculiar velocities used in the Pantheon+ analysis~\cite{Brout:2022vxf} inside balls defined by  the redshift limits specified in Table~\ref{t:nSn}. The result is shown in Fig.~\ref{f: vpec_distributions}. While in the mean these velocities are always negative, indicating infall, the statistical errors are large and $\bv_{\rm pec}\cd\bn =0$ is compatible within the statistical error, see left panel. Nevertheless, the mean radial peculiar velocities are by a factor $1.5$ to 5 smaller than our best fit value of $z_0$, see right panel. Due to the large errors in $\bv_{\rm pec}\cd\bn$ and in $z_0$, see also Fig.~\ref{f:z0_H0_contours}, the precise numbers should not be taken very seriously, but the trend is clear.

\begin{figure}[ht]
	\centering
	\includegraphics [scale=0.5]{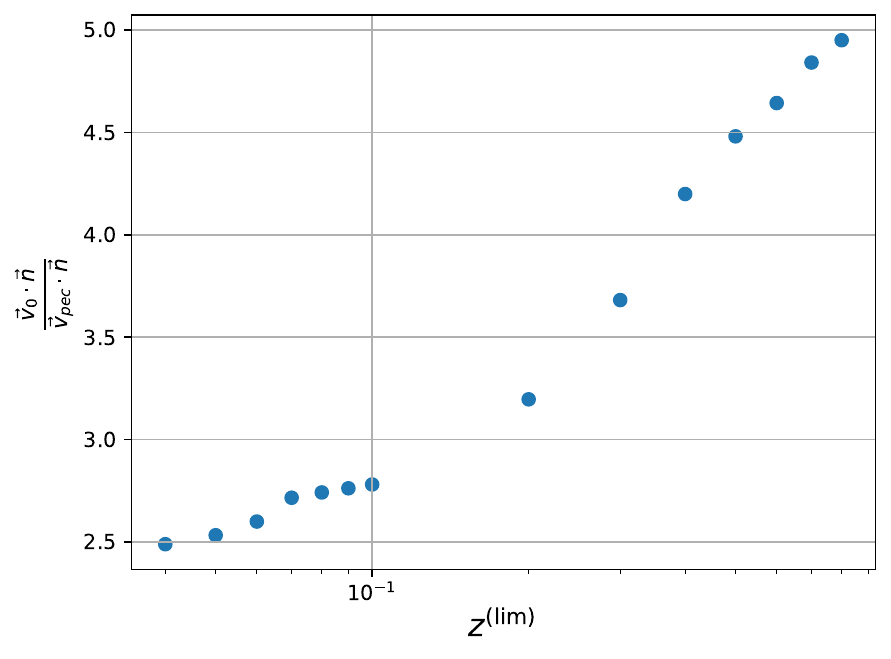}\includegraphics [scale=0.5]{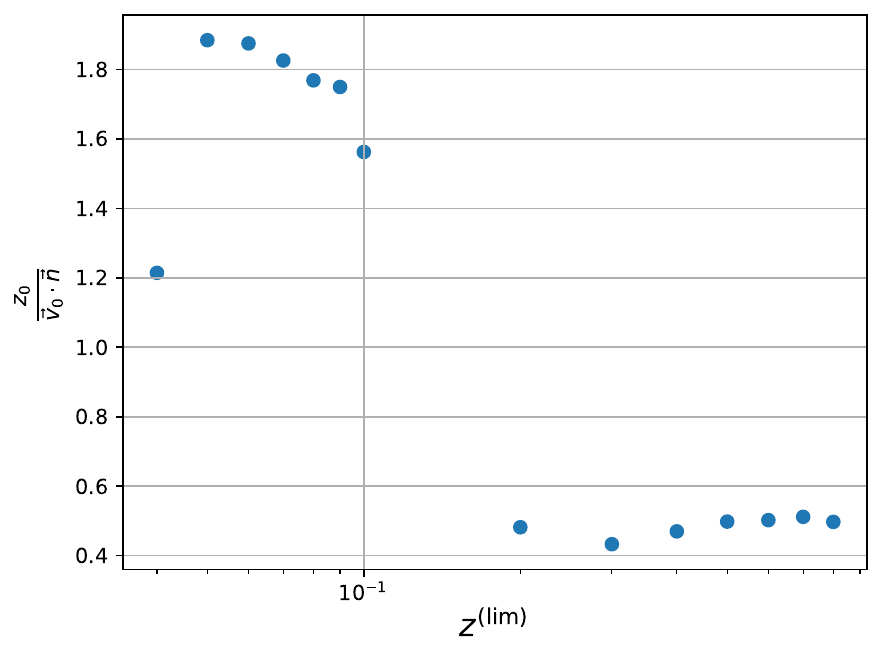}
\caption{Left panel: The agnostic bulk velocity is compared to the peculiar velocity as modelled in the original Pantheon+ analysis for different redshift limits.\\
Right panel: The radial infall velocity (in units where $c=1$) obtained for different redshift limits is compared with the corresponding bulk dipole.
\label{f: vpec_ratio}}
\end{figure}

If the supernovae were distributed uniformly over the sky, then the average of the bulk dipole over the supernova directions would vanish. However, for the actual distribution of supernovae, the average of $\bv_0 \cdot \bn$ is non-zero, and in the left panel of Fig.~\ref{f: vpec_ratio} we compare the direction-averaged bulk dipole to the radial part of the Pantheon+ peculiar velocities. Also the averaged bulk dipole velocity is 2.5 to 5 times larger in our agnostic analysis than the peculiar velocities in the Pantheon+ data.

Finally, comparing the best fit radial infall velocity $v_r=c z_0$ to the best fit bulk velocity $v_0$ (again averaged over supernova directions) we find that only for the lowest redshift limits, $z^{\rm (lim)}\leq0.1$, the preferred radial velocity is larger than the direction-averaged bulk velocity. However, as $z_0$ has large errors for $z^{\rm (lim)}\leq0.1$, see Fig.~\ref{f:z0_H0_contours}, also these numbers only indicate a trend. For $z^{\rm (lim)} \geq 0.2$ the best fit radial velocity is somewhat more than half of the bulk velocity, see right panel of Fig.~\ref{f: vpec_ratio}.  It is actually quite well approximated by $v_r=cz_0\sim v_0/\sqrt{3}$ as we would expect for a velocity component of a isotropic velocity field with three independent components.

\section{Conclusions}\label{s:con}
In this paper we analyse the Pantheon+ data by fitting an agnostic third order polynomial in redshift to the  luminosity distances. We find that in addition to the bulk velocity which we have already discussed in a previous work~\cite{Sorrenti_2023}, the data requires a radial infall velocity of about 100km/s for supernovae at redshift $z<0.04$, i.e.\ inside a ball of radius $R=0.04/H_0 =120h^{-1}$Mpc. Adding this infall improves the fit by $\De\chi^2>7$. 

In addition to the tests described in the previous section we performed two additional studies. As supernovae are independent events one may expect that the non-diagonal entries in the covariance matrix are to a large extent due to the modelling of correlated peculiar velocities which we do not use in our analysis as we work with the truly measured, uncorrected redshifts and fit corrections directly to the data. We therefore studied to what extent our findings depend on the non-diagonal entries of the covariance matrix. For this we have neglected the non diagonal entries in the covariance matrix and repeated parts of our analysis with only the diagonal entries in the covariance matrix. As one expects, we obtain tighter constraints in this case. However, the characteristics of our results, especially the degeneracies do not change.

Finally, we investigated what happens if we disregard the closest supernovae with $z<0.01$, following a similar test made in~\cite{peri}. This leaves us with only 7 remaining Cepheids and therefore significantly larger error in $dM$. This uncertainty of $dM$ leads to larger  errors also in the other parameters. But again, we do not see any significant trend in the results.

In Table~\ref{t:sum}, we summarize the comparison of different agnostic models and $\La$CDM with a dipole, including all supernovae up to $z^{\rm (lim)} = 0.8$ where our expansion starts to break down. 
\begin{table}[ht]
\centering
\begin{tabular}{|c|c|c|}
\hline
 Model & $\chi^2$ &  $\De\chi^2$ \\
 \hline
 agnostic with &&  \\
 dipole+monopole for $z\leq 0.04$ & 1475 & 
 --- \\
 \hline
 agnostic with &&  \\
 dipole+monopole for all SNe & 1475.4 &  0.44 
\\
\hline
 agnostic with &&  \\
 dipole & 1482.6 &  7.26 
\\
\hline
$\La$CDM with &&  \\
 dipole+monopole for all SNe & 1477.8 &  2.76
\\
\hline
\end{tabular}
\caption{We compare the $\La$CDM model with its best fit monopole and dipole with the agnostic model and report their $\De\chi^2$ with respect to the best fit model (top line).  \label{t:sum}}
\end{table}
The best fit is the agnostic model with dipole+monopole, but with a monopole only included for SNe with redshift $z\leq 0.04$. The difference to including a monopole for all SNe is however negligible. The difference to $\La$CDM is somewhat larger, but considering that $\La$CDM has one parameter less, this is also not very significant. Not including a monopole (third line), however clearly worsens the fit significantly. We note that for all choices of redshifts (heliocentric, CMB centric and HD) the dipole is highly significant. The total $\chi^2$ for the best model considering that it contains 1671 lightcurves from 1513 supernovae and has 8 model parameter is very reasonable. Our agnostic model, but also $\La$CDM with a dipole and a monopole are excellent  fits to this data.

Let us also estimate whether the  radial velocity corresponding to $z_0$ is compatible with the standard $\La$CDM cosmology.
We use linear perturbation theory as the scale of 
$120h^{-1}$Mpc is safely in the linear regime.
We use Planck parameters for the dimensionless curvature power spectrum~\cite{Planck:2018vyg},
\bea
 \frac9{25}\De_{\Psi}(k)=\De_{\zeta}(k) &=& \frac{1}{2\pi^2}k^3P_{\zeta}(k) = A_s\left(\frac kk_*\right)^{n_s-1}\,, \text{ where}\\
\log(10^{10}A_s) &=& 3.04\,, \quad n_s=0.965\,, \quad k_* =0.05{\rm Mpc}^{-1}\,. \nonumber
\eea
From the relation of $\zeta$ and the Bardeen potential in the matter era, $\Psi=\frac35\zeta$
we then obtain for the dimensionless velocity power spectrum,
see e.g.~\cite{Durrer:2020fza},
\be
\De_{v}(k) \simeq \left(\frac{H_0}{k}\right)^2\Om_m^{1.2}\De_{D}(k) \simeq \frac{4}{25}\left(\frac{k}{H_0}\right)^2\Om_m^{-0.8}A_s\left(\frac kk_*\right)^{n_s-1}
\ee
Inserting numbers for $k=(1/120)h/$Mpc we find
\be
\sqrt{\De_{v}((1/120)h/{\rm Mpc})} \simeq 7.7\times 10^{-4} \simeq 230{\rm km/s} \,.
\ee
Assuming $v_r$ to be one of three independent equally distributed Gaussian variables, we then expect to measure in a ball of radius 120$h^{-1}$Mpc a radial velocity
of about 
\be
v_r \sim v/\sqrt{3} \simeq 130 {\rm km/s} \,.
\ee
Hence the result of 96km/s which we found for our complete sample is very typical and close to  this average value. Note that also on the scale of $1000h^{-1}$Mpc the variance of the velocity field is still $\De_{v}(10^{-3}h/{\rm Mpc}) = (30{\rm km/s})^2$.
The velocity field remains considerable also on large scales.

We therefore conclude that apart from moving with a bulk velocity of about 317 km/s in direction (ra,dec) $= (204^o,-53^o)$, a ball of radius $R=120h^{-1}$Mpc around us is infalling with about 100km/s and hence we live in a mean {\em overdensity} of  $\sqrt{\De_{D}(120h/{\rm Mpc})} \simeq 0.022\ \approx 2\%$ on a scale of $120h^{-1}$Mpc. This is also in agreement with similar findings~\cite{Giani:2023aor, PASTEN2024101385}. The often discussed `local void' therefore is not at all realized on this scale, and we 
rather observe a local overdensity of somewhat more than 2\% inside a ball of radius 120$h^{-1}$Mpc.

\acknowledgments
We thank Camille Bonvin, Sveva Castello, Giulia Cusin, Nastassia Grimm, Davide Piras and Rick Watkins for interesting discussions. The authors acknowledge financial support from the Swiss National Science Foundation. The computations were performed at University of Geneva using the {\em Baobab} HPC service.

\vspace{1.5cm}

\bibliography{refs}

\providecommand{\href}[2]{#2}\begingroup\raggedright\begin{thebibliography}{10}

\bibitem{Bonvin:2005ps}
C.~Bonvin, R.~Durrer, and M.~A. Gasparini, {\it {Fluctuations of the luminosity distance}},  {\em Phys. Rev. D} {\bf 73} (2006) 023523, [\href{http://arxiv.org/abs/astro-ph/0511183}{{\tt astro-ph/0511183}}]. [Erratum: Phys.Rev.D 85, 029901 (2012)].

\bibitem{Brout:2022vxf}
D.~Brout et~al., {\it {The Pantheon+ Analysis: Cosmological Constraints}},  {\em Astrophys. J.} {\bf 938} (2022), no.~2 110, [\href{http://arxiv.org/abs/2202.04077}{{\tt arXiv:2202.04077}}].

\bibitem{Sorrenti_2023}
F.~Sorrenti, R.~Durrer, and M.~Kunz, {\it The dipole of the pantheon+sh0es data},  {\em Journal of Cosmology and Astroparticle Physics} {\bf 2023} (nov, 2023) 054.

\bibitem{Sorrenti:2024}
F.~Sorrenti, R.~Durrer, and M.~Kunz, {\it {The low multipoles in the Pantheon+SH0ES data}},  \href{http://arxiv.org/abs/2403.17741}{{\tt arXiv:2403.17741}}.

\bibitem{emcee}
D.~{Foreman-Mackey}, D.~W. {Hogg}, D.~{Lang}, and J.~{Goodman}, {\it {emcee: The MCMC Hammer}},  {\em Publ. Astron. Soc. Pac.} {\bf 125} (Mar., 2013) 306, [\href{http://arxiv.org/abs/1202.3665}{{\tt arXiv:1202.3665}}].

\bibitem{schwimmbad}
A.~M. Price-Whelan and D.~Foreman-Mackey, {\it schwimmbad: A uniform interface to parallel processing pools in python},  {\em The Journal of Open Source Software} {\bf 2} (sep, 2017).

\bibitem{autocorr}
J.~{Goodman} and J.~{Weare}, {\it {Ensemble samplers with affine invariance}},  {\em Communications in Applied Mathematics and Computational Science} {\bf 5} (Jan., 2010) 65--80.

\bibitem{getdist}
A.~Lewis, {\it Getdist: a python package for analysing monte carlo samples},  {\em ~} {\bf ~} (2019) [\href{http://arxiv.org/abs/1910.13970}{{\tt arXiv:1910.13970}}].

\bibitem{Planck:2018vyg}
{\bf Planck} Collaboration, N.~Aghanim et~al., {\it {Planck 2018 results. VI. Cosmological parameters}},  {\em Astron. Astrophys.} {\bf 641} (2020) A6, [\href{http://arxiv.org/abs/1807.06209}{{\tt arXiv:1807.06209}}]. [Erratum: Astron.Astrophys. 652, C4 (2021)].

\bibitem{Carr_redshift_pantheon+}
A.~Carr, T.~M. Davis, D.~Scolnic, K.~Said, D.~Brout, E.~R. Peterson, and R.~Kessler, {\it The pantheon+ analysis: Improving the redshifts and peculiar velocities of type ia supernovae used in cosmological analyses},  {\em Publications of the Astronomical Society of Australia} {\bf 39} (2022) e046.

\bibitem{Kogut:1993ag}
A.~Kogut et~al., {\it {Dipole anisotropy in the COBE DMR first year sky maps}},  {\em Astrophys. J.} {\bf 419} (1993) 1, [\href{http://arxiv.org/abs/astro-ph/9312056}{{\tt astro-ph/9312056}}].

\bibitem{Planck:2013kqc}
{\bf Planck} Collaboration, N.~Aghanim et~al., {\it {Planck 2013 results. XXVII. Doppler boosting of the CMB: Eppur si muove}},  {\em Astron. Astrophys.} {\bf 571} (2014) A27, [\href{http://arxiv.org/abs/1303.5087}{{\tt arXiv:1303.5087}}].

\bibitem{Planck:2018nkj}
{\bf Planck} Collaboration, N.~Aghanim et~al., {\it {Planck 2018 results. I. Overview and the cosmological legacy of Planck}},  {\em Astron. Astrophys.} {\bf 641} (2020) A1, [\href{http://arxiv.org/abs/1807.06205}{{\tt arXiv:1807.06205}}].

\bibitem{Saha:2021bay}
S.~Saha, S.~Shaikh, S.~Mukherjee, T.~Souradeep, and B.~D. Wandelt, {\it {Bayesian estimation of our local motion from the Planck-2018 CMB temperature map}},  {\em JCAP} {\bf 10} (2021) 072, [\href{http://arxiv.org/abs/2106.07666}{{\tt arXiv:2106.07666}}].

\bibitem{uncertainties}
E.~O. Lebigot, ``Uncertainties: a python package for calculations with uncertainties.'' \url{http://pythonhosted.org/uncertainties/}.

\bibitem{Lopes_2024}
M.~Lopes, A.~Bernui, C.~Franco, and F.~Avila, {\it Bulk flow motion detection in the local universe with pantheon+ type ia supernovae},  {\em The Astrophysical Journal} {\bf 967} (may, 2024) 47.

\bibitem{Shapely:2006}
D.~Proust, Q.~Hern\'an, E.~R. Carrasco, A.~Reisenegger, E.~Slezak, M.~Hern\'an, R.~D{\"u}nner, L.~Sodr\'e, M.~J. Drinkwater, Q.~A. Parker, and C.~J. Ragone, {\it The shapley supercluster: the largest matter concentration in the local universe},  {\em The Messenger (ESO)} {\bf 124} (06, 2006).

\bibitem{peri}
L.~Perivolaropoulos and F.~Skara, {\it {On the homogeneity of SnIa absolute magnitude in the Pantheon+ sample}},  {\em Monthly Notices of the Royal Astronomical Society} {\bf 520} (02, 2023) 5110--5125, [\href{http://arxiv.org/abs/https://academic.oup.com/mnras/article-pdf/520/4/5110/54034587/stad451.pdf}{{\tt https://academic.oup.com/mnras/article-pdf/520/4/5110/54034587/stad451.pdf}}].

\bibitem{Durrer:2020fza}
R.~Durrer, {\em {The Cosmic Microwave Background}}.
\newblock Cambridge University Press, 12, 2020.

\bibitem{Giani:2023aor}
L.~Giani, C.~Howlett, K.~Said, T.~Davis, and S.~Vagnozzi, {\it {An effective description of Laniakea: impact on cosmology and the local determination of the Hubble constant}},  {\em JCAP} {\bf 01} (2024) 071, [\href{http://arxiv.org/abs/2311.00215}{{\tt arXiv:2311.00215}}].

\bibitem{PASTEN2024101385}
E.~Pastén, S.~Galvez, and V.~H. Cárdenas, {\it Approximations for the divergence of the reconstructed local large-scale structure velocity field and its possible implications for cosmology},  {\em Physics of the Dark Universe} {\bf 43} (2024) 101385.

\end{thebibliography}\endgroup
\bibliographystyle{JHEP}

\end{document}